\documentclass[12pt]{article}
\usepackage{amssymb,amsfonts}
\usepackage{amscd}
\newcommand{\be}{\begin{equation}}
\newcommand{\ee}{\end{equation}}

\newcommand{\fg}{\mathfrak{g}}
\newcommand{\al}{\alpha}

\newcommand{\bet}{\beta}
\newcommand{\g}{\gamma}
\newcommand{\Om}{\Omega}
\newcommand{\om}{\omega}
\newcommand{\lr}{\longleftrightarrow}

\newcommand{\G}{\Gamma}

\newcommand{\La}{\Lambda}

\newcommand{\M}{{\cal M}}
\newcommand{\Pen}{{\cal P}}

\newcommand{\by}{\bar{y}}

\newcommand{\la}{\lambda}

\newcommand{\eps}{\varepsilon}
\newcommand{\bz}{\bar{z}}

\newcommand{\w}{\omega}

\newcommand{\pn}{\tilde{\cal P}}
\newcommand{\dz}{\wedge}

\newcommand{\ba}{\begin{array}}
\newcommand{\ea}{\end{array}}
\newcommand{\beq}{\begin{eqnarray}}
\newcommand{\eeq}{\end{eqnarray}}
\newtheorem{lm}{Lemma}
\newtheorem{theor}{Theorem}
\newtheorem{pr}{Proposition}
\newtheorem{co}{Corollary}
\newtheorem{exa}{Example}
\newtheorem{deff}{Definition}
\newcommand{\bd}{\begin{deff}}
\newcommand{\ed}{\end{deff}}
\newcommand{\bl}{\begin{lm}}
\newcommand{\el}{\end{lm}}
\newcommand{\bp}{\begin{pr}}
\newcommand{\ep}{\end{pr}}
\newcommand{\bt}{\begin{theor}}
\newcommand{\et}{\end{theor}}
\newcommand{\bc}{\begin{co}}
\newcommand{\ec}{\end{co}}
\newcommand{\bex}{\begin{exa}}
\newcommand{\eex}{\end{exa}}

\newcommand{\der}{{\rm d}}
\begin{document}

\thispagestyle{empty}

\title{EXTENSIONS OF BUNDLES OF NULL DIRECTIONS
\footnote{Research supported in part by 
Komitet Bada\'n Naukowych Grant nr 2 P302 112 7, EPSRC 
grant GR/L65925, Erwin Sch\"odinger International Institute and Consorzio per 
lo Sviluppo Internazionale dell Universita degli Studi di Trieste.}\\ 
\vskip 1.truecm
{\small {\sc Pawe\l  ~Nurowski}}
\footnote{Instytut Fizyki Teoretycznej, 
Uniwersytet Warszawski, Warszawa, Poland, and Dipartimento di 
Science Mathematiche, Universita degli Studi di Trieste, Trieste, Italy. 
E-mail: nurowski@fuw.edu.pl}\quad 
{\small {\sc Lane Hughston}}\footnote
{Merill Lynch International, 25 Ropemaker Street, London EC2Y 9LY, UK. E-mail: lane@ml.com}\quad 
{\small {\sc David Robinson}}\footnote{E-mail: david.c.robinson@kcl.ac.uk}\\
{\small 
{\it Department of Mathematics,  King's College London, Strand, London WC2R 
2LS, UK}}\\
\vskip -0.3truecm
}
\author{\mbox{}}
\maketitle
\begin{abstract}
\noindent
The geometry of $\Pen$, the bundle of null directions over an Einstein 
space-time, is studied. 
The full set of invariants of the  natural $G$-structure on $\Pen$ is constructed using the 
Cartan method of equivalence. This leads to an extension of $\Pen$ which is an elliptic 
fibration over the space-time. Examples are given which show that such an 
extension, although natural, is not unique. A reinterpretation of the 
Petrov classification in terms of the fibres of an extension of $\Pen$ is 
presented.   
\end{abstract}

\newpage
\noindent
\section{Introduction}
In 1922 Elie Cartan made the following observation\cite{bi:car}.\\

\noindent
{\it From a geometric viewpoint, it is worthwhile to note an interesting property. At 
each point $A$ [of a conformally non-flat space-time] there exist four privileged 
null directions [...]. They can be 
characterized as follows: Any one of these directions, say $AA'$, is invariant under 
transport around an infinitesimal parallelogram one of whose sides is $AA'$ and the 
other of whose sides is along an arbitrary null direction at $A$. In the case of 
$ds^2$ corresponding to a single attractive mass ($ds^2$ of Schwarzschild) the four 
privileged directions reduce to two (degenerate) directions which correspond to null 
rays pointing to or from the center of attraction.}\footnote{Translation of A. 
Magnon and A. Ashtekar from Ref. \cite{bi:ab}. We thank A. Trautman for 
informing us about this quotation.}\\

\noindent
This remark implicitly anticipates elements of the so called Petrov 
classification, the elegant 
contemporary formulation of which owes much to the work of Roger Penrose. In this 
formulation the anti-selfdual part of the Weyl tensor at a space-time point corresponds to a totally 
symmetric spinor $C_{ABCD}$ and a null direction at a point is defined uniquely 
in terms of a spinor $\xi^A(z)=\left(\begin{array}{c}1\\z\end{array}\right)$. Then 
the Cartan (principal) null directions correspond to the solutions of the following 
equation 
\be
C_{ABCD}\xi^A\xi^B\xi^C\xi^D=0,
\label{eq:cpp}
\ee
for $z\in{\bf C}\cup\{\infty\}$ defining the spinor $\xi^A$. This equation 
being fourth order in $z$ always has four roots, $z_i$ say, but some of them 
may be repeated. Multiple roots correspond to coincidences between the 
corresponding Cartan  null directions. The Petrov classification (or 
the Cartan-Petrov-Penrose classification \cite{bi:car,bi:pet,bi:pen} 
as it perhaps should be properly called) of metrics at a given space-time 
point is based on these results. One says that the metric is algebraically 
general at a point if its Weyl tensor defines four distinct Cartan directions 
there. Otherwise the metric is algebraically special. The following five 
possibilities 
may occur.\\

\noindent
\centerline{$z_1$, $z_2$, $z_3$, $z_4$ all different $\lr$ 4 distinct Cartan directions $\lr$ Petrov 
type I}
\centerline{$z_1=z_2$, $z_3, z_4$ different $\lr$ 3 distinct Cartan 
directions $\lr$ Petrov type II}
\centerline{$z_1=z_2\neq z_3=z_4$ different $\lr$ 2 pairs of  distinct Cartan 
directions $\lr$ Petrov type D}
\centerline{$z_1=z_2=z_3$, $z_4$ different $\lr$ 2 distinct Cartan 
directions $\lr$ Petrov type III}
\centerline{$z_1=z_2=z_3=z_4$ $\lr$ 1  Cartan direction $\lr$ Petrov type N}

\mbox{}\\
\noindent
This classification can be reinterpreted as follows.  Consider a curve 
\be
w^2=C_{ABCD}\xi^A\xi^B\xi^C\xi^D\label{eq:cpp1}
\ee
in ${\bf C}^2$ with coordinates $(w,z)$ or, better, a compact 
Riemann surface $\cal T$ 
associated with a double valued function $w(z)=\sqrt{C_{ABCD}\xi^A\xi^B\xi^C\xi^D}$ on 
${\bf C}\cup\{\infty\}\cong{\bf S}^2$. It is well known that the topology of $\cal T$ 
depends on the roots of (\ref{eq:cpp}) and corresponds to a 2-dimensional torus 
${\bf T}^2$ if all $z_i$ are distinct. If some coincidences between the roots occur 
then we have the following possibilities. $\cal T$ has the topology of a torus with one 
vanishing cycle in Petrov type II, it has the topology of two spheres touching each other in two different 
points in Petrov type D, it has the topology of a sphere with a distinguished point in Petrov type III,  
and it has the topology of two 
spheres touching each other in a single point in Petrov type N.\\

\noindent
It turns out that equation (\ref{eq:cpp1}), which seems to be artificially added, 
appears naturally in the Einstein theory \cite{bi:formul,bi:einstein}. A fibration 
$\pn$ can be defined over the space-time, each fibre having the topology of the associated surface 
$\cal T$, with the Einstein equations taking an interesting form on the total space 
\cite{bi:formul,bi:einstein}. In this paper we extend 
the results of Refs. \cite{bi:formul,bi:einstein} by showing how the fibration $\pn$ 
can be defined by using natural objects on the Penrose bundle of null 
directions $\Pen$ over the space-time.\\

\noindent
We recall that given a 4-dimensional Lorentzian manifold $({\cal M},g)$ and its 
bundle of null directions $\Pen$ one naturally defines a class of six 1-forms 
$[(F,\bar{F},T,\La,E,\bar{E})]$ on it having the following properties \cite{bi:optical,bi:traut} 
(see also Section 5 of the present paper):
\begin{itemize}
\item[({\it i.})]
$\La$, $T$ are real- and $F$, $E$ are complex-valued 1-forms on $\Pen$
\item[({\it ii.})]
$F\dz\bar{F}\dz T\dz \La\dz E\dz\bar{E}\neq 0$ at each point $p$ of 
$\Pen$
\item[({\it iii.})]
Two sets of forms 
$(F, \bar{F}, T,\La , E,\bar{E})$ and 
$(F', \bar{F}', T',\La', E',\bar{E}')$ are in the same class iff 
\be
\Lambda=\frac{1}{A}\Lambda',
\ee
\be
F={\rm e}^{i\varphi}( F' +\by \Lambda')
\ee
\be
T=A(T'+\by\bar{F}'+yF'+y\by\Lambda' )
\ee
\be
E=\frac{1}{w} E',
\ee
\end{itemize}
where $A>0$, $\varphi$ (real) $y$, $w\neq 0$ (complex) are 
arbitrary functions on $\Pen$. This defines a certain $G$-structure on $\Pen$. This 
structure can be studied using the Cartan method of equivalence. In this paper we solve the Cartan 
equivalence problem for this $G$-structure. We show that this naturally leads to an elliptic fibration 
$\pn$ associated with the Einstein space-time i.e. to the  association of an elliptic curve 
(\ref{eq:cpp1}) with each point of the space-time. The extension of $\Pen$ to $\pn$ was obtained 
previously in Refs. \cite{bi:formul,bi:einstein} by the continuation of solutions of a certain 
differential system $\cal I$, defined initially only on an open set of ${\bf R}^6$. Such an extension 
is natural but, as is discussed below, not unique.\\

\noindent
The paper is organized as follows. Section 2 contains notation and definitions. 
Section 3 defines the differential system $\cal I$. Section 4 gives examples of 
solutions of the differential system $\cal I$ corresponding to all vacuums of type N. 
Section 5 uses the Cartan method of equivalence to get the differential 
system of Section 4 from the natural objects defined on the bundle of null 
directions over the space-time. The results of Secion 5 are applied in Section 6 to 
give an effective algorithm for checking whether two metrics are isometrically 
equivalent. In Section 7 a way of associating an elliptic curve with any
point of a conformally non-flat Einstein space-time is presented. The elliptic fibration 
associated in this way with any conformally non-flat Einstein 
space-time constitutes a double branch cover of the bundle of null directions. 
This natural extension of $\Pen$ is not unique and in 
Section 8 some examples of Einstein space-times, with different extension of 
$\Pen$, are exhibited.    
\section{Basic definitions}

\noindent
We briefly recall the definitions of the geometrical 
objects we need in the following. Let $\M$ be a 4-dimensional 
oriented and time-oriented 
manifold equipped with a Lorentzian metric $g$ of signature $(+,+,+,-)$. 
It is convenient to introduce a null frame $(m,~\bar{m},~k,~l)$ on $\M$ 
with a dual coframe $\theta^i=(\theta^1,~\theta^2,~\theta^3,~\theta^4)=
(M,~\bar{M},~K,~L)$ so that
\footnote{Expressions such as $\theta^i\theta^j$ mean the symmetrized tensor 
product, e.g. $\theta^i\theta^j=\frac{1}{2}(\theta^i\otimes\theta^j+ 
\theta^j\otimes\theta^i)$. Also, we will denote by round (resp. square) 
brackets the symmetrization (resp. antisymmetrization) of indices, e.g. 
$a_{(ik)}=\frac{1}{2}(a_{ik}+a_{ki})$, $a_{[ik]}=
\frac{1}{2}(a_{ik}-a_{ki})$, etc.}
\be 
g=g_{ij}\theta^i\theta^j=2(M\bar{M}-KL).
\label{eq:met} 
\ee 
%Here $\theta^i\theta^j=\theta^i\otimes \theta^j+\theta^j\otimes \theta^i$, 
%etc. 
Given $g$ and $\theta^i$ the connection 1-forms 
$\G_{ij}=g_{ik}\G^k_{~j}$ are uniquely defined by 
\be
\der\theta^i=-\G^i_{~j}\dz\theta^j,~~~~~~~\G_{ij}+\G_{ji}=0.\label{eq:kon}
\ee
The connection coefficients $\G_{ijk}$ are determined 
by the relation $\G_{ij}=\G_{ijk}\theta^k$.\footnote{We lower and raise indices by means 
of the metric and its inverse.} 
Using them we define the curvature 2-forms ${\cal R}^k_{~i}$, the 
Riemann tensor $R^{i}_{~jkl}$, the Ricci tensor 
$R_{ij}$ and the Ricci scalar $R$ by 
$$
{\cal R}^k_{~i}=\frac{1}{2}R^k_{~imj}\theta^m\dz\theta^j=\der\G^k_{~i}+
\G^k_{~j}\dz\G^j_{~i},~~~R_{ij}=R^k_{~ikj},~~~R=g^{ij}R_{ij}.
$$
We also introduce the traceless Ricci tensor by 
$$
S_{ij}=R_{ij}-\frac{1}{4}g_{ij}R.
$$
Note that the vanishing of $S_{ij}$ is equivalent to the Einstein equations 
$R_{ij}=\la g_{ij}$ for the metric $g$. We define the Weyl tensor 
$C^i_{~jkl}$ by
$$
C_{ijkl}=R_{ijkl}+\frac{1}{3}Rg_{i[k}g_{l]j}+R_{j[k}g_{l]i}+R_{i[l}g_{k]j},
$$
and its spinorial coefficients $\Psi_\mu$ by
%\footnote{Theses equations are copied from Ref. \cite{bi:nur1}. 
%There, the curvature 2-forms ${\cal R}_{ij}$ 
%were denoted by  $\Om_{[ij]}$. 
%We take this opportunity to mention that a sign missprint was present 
%in the expressions for $\Om_{[23]}$ and $\Om_{[13]}$ in appendix B of 
%\cite{bi:nur1}. The correct 
%formulae (B1) and (B2) of \cite{bi:nur1} should have minus sign in front of 
%$\frac{1}{12}R$ in the expressions for 
%$\Om_{[23]}$ and $\Om_{[13]}$. Note also 
%that because of the orientation change, formulae of \cite{bi:nur1} 
%translate to the present paper by respecting the following rules. 
%(a) Put $N$ and $P$ of \cite{bi:nur1} to be equal to 
%the present $L$ and $\bar{M}$, respectively. 
%(b) Replace all the primes in \cite{bi:nur1} by bars, 
%e.g. $\Psi_2'\to\bar{\Psi}_2$, $z'\to\bar{z}$, etc.  
%(c) Interchange the indices 3 and 4 in all of the formulae.}
\beq                 
&{\cal R}_{23}=\Psi_4 \bar{M}\dz K+\Psi_3 (L\dz K-M\dz \bar{M})+
(\Psi_2 +\frac{1}{12}R)L\dz M\nonumber\\
&+\frac{1}{2}S_{33}M\dz K+\frac{1}{2}S_{32}(L\dz K+M\dz\bar{M})+
\frac{1}{2}S_{22}L\dz\bar{M},\nonumber\\
&\quad\nonumber\\
&{\cal R}_{14}=(-\Psi_2-\frac{1}{12}R)\bar{M}\dz K-
\Psi_1(L\dz K-M\dz \bar{M})-
\Psi_0 L\dz M\nonumber\\
&-\frac{1}{2}S_{11}M\dz K-\frac{1}{2}S_{41}(L\dz K+M\dz\bar{M})-
\frac{1}{2}S_{44}L\dz\bar{M},\label{eq:r23}\\
&\quad\nonumber\\
&\frac{1}{2}({\cal R}_{43}-{\cal R}_{12})=\Psi_3 \bar{M}\dz K+
(\Psi_2-\frac{1}{24}R) 
(L\dz K-M\dz \bar{M})+\Psi_1 L\dz M\nonumber\\
&+\frac{1}{2}S_{31}M\dz K+\frac{1}{4}(S_{12}+S_{34})(L\dz K+M\dz\bar{M})+
\frac{1}{2}S_{42}L\dz\bar{M}.\nonumber
\eeq~\\

\noindent
\section{The differential system}

\noindent
Here we quote the major results from Ref. \cite{bi:formul} which will be used 
in this paper. They describe the properties of a system of six 1-forms on 
${\bf R}^6$which determine a conformally non-flat Lorentzian 4-metric 
satisfying Einstein equations.
\bt~\\ 
Let $\Pen_0$ be an open subset of ${\bf R}^6$. Suppose that on 
$\Pen_0$ we have six 1-forms $(F, \bar{F}, T,\La, E,\bar{E})$ which 
satisfy the following conditions 
\begin{itemize}
\item[{\bf i)}]
$T,\La$ are real- and $F,E$ are complex-valued 1-forms
\item[{\bf ii)}] 
$F\dz\bar{F}\dz T\dz\La\dz E\dz\bar{E}\neq 0$ at each point $p$ of $\Pen_0$
\item[{\bf iii)}] there exist complex-valued 1-forms $\Om$ and $\G$ on 
$\Pen_0$, and a certain complex function $\al$ on $\Pen_0$ such that 
$$
{\rm d}F=(\Om -\bar{\Om})\dz F+E\dz T+\bar{\G}\dz\La
$$
\be
{\rm d}T=\G\dz F + \bar{\G}\dz\bar{F}-(\Om +\bar{\Om})\dz T\label{eq:sys1}
\ee
$$
{\rm d}\La=\bar{E}\dz F+ E\dz\bar{F}+(\Omega+\bar{\Omega})\dz\La
$$
$$
{\rm d}E=2\Om\dz E+\bar{F}\dz T+\al \La\dz F.
$$
\end{itemize}
Then\\
\begin{itemize} 
\item[i)] 
$\Pen_0$ is locally foliated by 2-dimensional manifolds ${\cal S}_x$, which 
are tangent to the real distribution $\cal V$ defined by
$$
F({\cal V})=T({\cal V})=\La ({\cal V})=0,
$$
\item[ii)]
the degenerate metric 
\be
G=2(F\bar{F}-T\La) 
\ee
on $\Pen_0$ has the signature (+,+,+,--,0,0) and is preserved when Lie-transported along any 
leaf ${\cal S}_x$ of the foliation $\{{\cal S}_x\}$,
\item[iii)] 
the 4-dimensional space $\cal M$ of 
all leaves of the foliation $\{{\cal S}_x\}$ is naturally equipped with a 
Lorentzian conformally non-flat metric $g$ which is Einstein 
($S_{ij}=0$) and is defined by projecting $G$ from $\Pen_0$ to 
$\cal M$.
\end{itemize}
\et

\noindent
The Einstein property of the metric $g$ was proven in \cite{bi:formul} 
by using the integrability conditions for the system (\ref{eq:sys1}). We  
summarize them in the following proposition. 

\bp~\\
If $(F, \bar{F},T,\La,E,\bar{E})$ satisfy (\ref{eq:sys1}) then there 
exist complex functions $a,h$ on $\Pen_0$, and a real constant $\la$, such 
that  
$$
{\rm d}\G=2\G\dz\Om+\al T\dz \bar{F}+a(T\dz\La+
F\dz\bar{F})+h\La\dz F
$$
\be
{\rm d}\Om=E\dz\G-(\al+\frac{\la}{2})(T\dz\La+F\dz\bar{F})+a\La\dz F.
\label{eq:sys2}
\ee
Moreover,
\beq
&\der\al=\al_1F+\g_4\bar{F}+\g_1T+\al_4\La-2aE,\nonumber\\
&\der a=a_1F+\al_4\bar{F}+\al_1T+a_4\La+hE-(3\al+\la)\G-2a\Om,
\label{eq:domy1}\\
&\der h=h_1F-a_4\bar{F}-a_1T+h_4\La+4a\G-4h\Om,\nonumber
\eeq
where the possible forms of $\Om$ and $\G$ are 
\beq
&\Om=\om_1F+\om_2\bar{F}+\om_3T+\om_4\La,\label{eq:domy2}\\
&\G=\g_1F-4\om_4\bar{F}-4\om_1T+\g_4\La-(3\al+\la)E \nonumber
\eeq
and $\g_1,\g_4,\al_1,\al_4,a_1,a_4,h_1,h_4,\om_1,\om_2,\om_3,\om_4$ are 
certain complex functions on $\Pen_0$.\\
\ep

\noindent
The next result gives a geometric interpretation of the functions $\al$, $h$ and 
the constant $\la$.

\bp~\\
The spinorial coefficients for the Weyl 
tensor of the metric $g$ on $\cal M$ read   
%These, in the tetrad $(F,\bar{F},T,\La)$, 
$$
\Psi_0=h,~~~\Psi_1=-a,~~~\Psi_2=-\al -\frac{\la}{3},~~~
\Psi_3=0,~~~\Psi_4=-1.
$$
The metric is of  Petrov   
type D iff 
\be
a=0, ~~~~~~{\rm and}~~~~~ h=-(3\al+\la)^2 \label{eq:D}
\ee
and of type N iff
\be
h=a=0,~~~~~~{\rm and}~~~~~\la=-3\al.\label{eq:N}
\ee
The metric is algebraically special iff $I^3=27J^2$, where 
$$
I=-h+\frac{1}{3}(3\al+\la)^2, ~~~{\rm and}~~~J=a^2+
\frac{1}{27}(3\al+\la)^3+\frac{1}{3}h(3\al+\la).
$$
\ep

\noindent 
From now on the differential system $\cal I$ of the forms 
$(F,\bar{F},T,\La,E,\bar{E})$ satisfying equations $(\ref{eq:sys1})$ on 
$\Pen_0$ will be denoted by $({\cal I},\Pen_0)$.

\noindent
\section{Type N vacuum solutions}

\noindent
In this section we present some specific examples of the differential systems 
$({\cal I},\Pen_0)$. We exhibit the forms $(F,\bar{F},T,\La,E\bar{E})$ 
which, via Theorem 1, correspond to type N vacuum 
$(\al=\la=0)$ solutions of the Einstein equations. These examples illustrate the way in which well known solutions appear in this formalism.\\ 

\noindent
{\bf Example 1}\\
Let $\Pen_0$ be an open set of ${\bf R}^6$ with coordinates 
$(Z,\bar{Z},U,V,z,\bz)$, where $U$, $V$ are real and $Z$, $z$ are 
complex. Consider the following 1-forms on $\Pen_0$.
$$
F=\der \bar{Z}+z\der U
$$
\be
T=\der U\label{eq:ex1}
\ee
$$
\La=\der V +z\der Z +\bz\der\bar{Z}+[z\bz -\frac{1}{2}(Z^2+\bar{Z}^2)]\der U
$$
$$E=\der z + Z\der U.$$

It is a matter of straightforward calculation to check that these forms 
constitute a solution to the system (\ref{eq:sys1}) with $\al=\la =\Om=\G=0$.

One also easily checks that although $F,T,\La, E$ depend on six real 
coordinates, the metric 
$$
G=2(F\bar{F}-T\La)=
2\der Z\der \bar{Z}-2\der U[\der V-\frac{1}{2}(Z^2+\bar{Z}^2)\der U]
$$
depends on $(Z,\bar{Z},U,V)$ only. Thus, it projects 
to 
$$g=2\der Z\der \bar{Z}-2\der U[\der V-\frac{1}{2}(Z^2+\bar{Z}^2)\der U]$$
on a 4-manifold $\cal M$ coordinatized by $(Z,\bar{Z},U,V)$. The space-time 
$\cal M$ with this metric is a plane-fronted gravitational wave possessing 
6 symmetries.\\ 

\noindent
{\bf Example 2}\\
Let $\Pen_0$ be again coordinatized by $(Z,\bar{Z},U,V,z,\bz)$ 
and let $\om_3$=const be a complex parameter. Consider the forms  
$$
F=\der \bar{Z}+[z+(\bar{\om}_3-\om_3)\bar{Z}]\der U
$$
\be
T=\der U\label{eq:ex2}
\ee
$$
\La=\der V +z\der Z +\bz\der\bar{Z}+[z\bz -\frac{1}{2}(Z^2+\bar{Z}^2)+
(\om_3-\bar{\om}_3)(zZ-\bz\bar{Z})-(\om_3+\bar{\om}_3)v]\der U
$$
$$E=\der z + (Z-2\om_3 z)\der U.$$

\noindent
They again constitute a solution to the system (\ref{eq:sys1}). This solution 
generalizes the previous example since the corresponding $\al=\la=\G=0$, but  
$\Om=\om_3T$. For any value of the complex parameter $\om_3$ the metric 
$$
g=2\der Z\der \bar{Z}-2\der U[\der V+(\om_3-\bar{\om}_3)(\bar{Z}\der Z-
Z\der\bar{Z})+((\om_3-\bar{\om}_3)^2Z\bar{Z}-(\om_3+\bar{\om}_3)V-
\frac{1}{2}(Z^2+\bar{Z}^2))\der U]
$$  
on the quotient manifold parametrized by $(Z,\bar{Z},U,V)$ is a 
plane-fronted gravitational wave with 6 symmetries. The solution of Example 1 
corresponds to $\om_3=0$ and is the simplest in the class. It follows that 
Example 2 exhausts the list of all vacuum ($\la=0$) solutions to the Einstein 
equations with 6 symmetries.\\

\noindent
{\bf Example 3}\\
A generalization of the preceeding examples can be obtained by taking 
$\Pen_0$ with coordinates $(Z,\bar{Z},U,V,z,\bz)$ and the forms 
$$
F={\rm e}^{-i\phi}[\der \bar{Z}+z\der U]
$$
\be
T={\rm e}^r\der U\label{eq:ex3}
\ee
$$
\La={\rm e}^{-r}[\der V +z\der Z +\bz\der\bar{Z}+(z\bz -H-\bar{H})\der U]
$$
$$E={\rm e}^{-r-i\phi}[\der z + H_Z\der U].$$
Here $H=H(Z,U)$ is any holomorphic function of the variable $Z$, 
$H_Z=\partial H/\partial Z$ and the real 
functions $r$ and $\phi$ are determined by the condition 
${\rm e}^{2(r+i\phi)}=H_{ZZ}$. One easily sees that equations (\ref{eq:ex3}) 
constitute a solution to the system (\ref{eq:sys1}) with $\al=\la=\G=0$ 
and $\Om=-\frac{1}{2}\der (r+i\phi)$. The corresponding type N vacuum 
space-time is a general plane wave with 5 (or more) symmetries.\\

\noindent 
{\bf Example 4}\\
Another example of solutions with $\al=\la=0$ is given by  
$$
F={\rm e}^{-i\phi}[\der \bar{Z}+(V+z(Z+\bar{Z}))\der U]
$$
\be
T={\rm e}^r(Z+\bar{Z})\der U\label{eq:ex4}
\ee
$$
\La={\rm e}^{-r}[\der V +z\der Z +\bz\der\bar{Z}+(z\bz(Z+\bar{Z})+(z+\bz)V
-H-\bar{H})\der U]
$$
$$E={\rm e}^{-r-i\phi}[\der z + (z^2+H_Z)\der U].$$
Here $\Pen_0$ is parametrized by $(Z,\bar{Z},U,V,z,\bz)$, $H=H(Z,U)$ is 
holomorphic in $Z$ and the real functions $r$ and $\phi$ are given by 
${\rm e}^{2(r+i\phi)}=H_{ZZ}/(Z+\bar{Z})$. The solutions are type N vacuums 
($\al=\la=0$) and have $\Om=-\frac{1}{2}\der (r+i\phi)-z\der U$ and 
$\G=-{\rm e}^{r+i\phi}\der U$. They belong to the Kundt class.\\

\noindent
All the solutions presented so far corresponded to type N vacuums with 
nondiverging rays. Generic type N vacuums are given 
below.\\

\noindent
{\bf Example 5}\\
Consider $\Pen_0\subset {\bf R}^6$ with coordinates 
$(Z,\bar{Z},U,V,z,\bz)$, where $Z,z$ are complex and $U,V$ are real. 
Define
$$
\La_0=\der U+\xi\der Z+\bar{\xi}\der\bar{Z},
$$
where $\xi=\xi(U,Z,\bar{Z})$ is a function of variables $U,Z,\bar{Z}$ 
only. Let  
$$
F={\rm e}^{-i\phi}
[\der \bar{\xi}+V\der Z-\bar{\partial}\bar{\xi}\der\bar{Z}+z\La_0]
$$
\be
T={\rm e}^r\La_0\label{eq:ex5}
\ee
$$
\La={\rm e}^{-r}[\der V+z\der \xi +\bz\der\bar{\xi}+z\bz\La_0+
(\partial\bar{\partial}\bar{\xi}+zV-\bz\bar{\partial}\bar{\xi})\der\bar{Z}+
(\bar{\partial}\partial \xi+\bz V-z\partial \xi)\der Z]
$$
$$
E={\rm e}^{-r-i\phi}[\der z + (z^2+\partial_U\bar{\partial}\bar{\xi})\der
\bar{Z}],
$$
where $\partial_U=\partial/\partial U$, $\partial=\partial_Z-\xi\partial_U$ 
and real functions $r$ and $\phi$ are determined by 
${\rm e}^{2(r+i\phi)}=
-\partial_U^{~2}\bar{\partial}\bar{\xi}/(V+\bar{\partial}\xi)$. It follows 
that 
if the function $\xi$ satisfies the equations
$$\partial\partial_U\bar{\partial}\bar{\xi}=0$$
$${\rm Im}\partial\partial\bar{\partial}\bar{\xi}=0$$
then the above forms satisfy the system (\ref{eq:sys1}) with
$\al=\la=0$ and $\Om=-\frac{1}{2}\der (r+i\phi)-z\der \bar{Z}$ and 
$\G=-{\rm e}^{r+i\phi}\der Z$. The corresponding space-times are of type N.\\

\noindent
It follows from the results of Pleba\'nski \cite{bi:Pleban} that 
Examples 3, 4, 5 constitute all the solutions to the vacuum type N Einstein 
equations.\\

\noindent
\section{From the Einstein space-time to the differential system}

\noindent
We first briefly summarize the Cartan method of analyzing $G$-structures (see 
Ref. \cite{bi:Jac} for more details).\\

\noindent
Let $\cal X$ be a $n$-dimensional manifold and $[\{\theta^i\}]$, $i=1,...,n,$ be a 
class of $n$ linearly independent 1-forms on $\cal X$ such that two representatives 
$\{\theta^i\}$ and $\{\theta'^i\}$ are in the same class iff there exists an element 
$(a^i_j)\in G$ of a certain group $G$ such that  $\theta'^i=a^i_j\theta^j$. Now, 
suppose that we have two sets $\{\theta'^i\}$ and $\{\theta^i\}$ of $n$ linearly 
independent 1-forms on $\cal X$. The $G$-structure equivalence question is: does there exists a 
(local) 
diffeomorphism $\phi$ of $\cal X$ such that 
\be
\phi^*(\theta'^i)=a^i_j\theta^j\label{eq:roo}
\ee
for some 
$G$-valued function $a^i_j$ on $\cal X$. In other words does the system of differential equations 
(\ref{eq:roo}), for $\phi$, have a 
solution. This question is not easy to answer, since the right hand side of (\ref{eq:roo}) is 
undetermined. Elie Cartan associates with the forms $\{\theta^i\}$ and $\{\theta'^i\}$ two  
systems of 1-forms ${\Om_\mu}$ and ${\Om'_\mu}$ on a manifold $\hat{\cal X}$ of 
dimension $\hat{n}\geq n$. Then he shows that equations like (\ref{eq:roo}) for $\phi$ 
have a solution iff a 
simpler system 
\be
\hat{\phi}^*\Om'_\mu=\Om_\mu\label{eq:roo1}
\ee
 of differential equations for a 
diffeomorphism $\hat{\phi}$ of $\hat{\cal X}$ has a solution. Examples are known 
(e.g. CR-structures \cite{bi:carcr}) where the Cartan procedure produces $k>\hat{n}$ 
1-forms $\Om_\mu$ of which $\hat{n}$ are linearly independent. Then decomposing 
$k-\hat{n}$ of the dependent 1-forms $\Om_\mu$ onto the basis of $\hat{n}$ independent ones 
we get functions $f_I$ (coefficients of the decompositions) which if (\ref{eq:roo1}) 
has a solution have to satisfy $\hat{\phi}^*(f'_I)=f_I$. The advantage of these 
equations for $\hat{\phi}$ is that they are not differential equations. 
If the procedure gives 
enough independent functions $f_I$ then by the implicit function theorem 
the whole problem reduces to evaluating whether a certain Jacobian is non-degenerate. \\

\noindent
In this Section we show that any conformally non-flat Einstein 
space-time defines a differential system as in Theorem 1. We consider 
the bundle $\cal P$ of null directions of an Einstein space-time and study 
its natural $G$-structure using the Cartan 
%\footnote{This structure was described by one of us in 
%Refs. \cite{bi:optical,bi:traut}. In this paper it is defined as the 
%class of 1-forms satisfying conditions $(i.)-(iii.)$ of this Section.}. 
method described above. This enables us to define a differential system on $\cal P$ that has all 
the properties of $({\cal I}, \Pen_0)$. Here the arguments presented  in Ref. \cite{bi:einstein} 
are approached from a different point of view.\\

\noindent
Let $({\cal M}, g)$ be a 4-dimensional Lorentzian (not necessarily Einstein) 
manifold.
Consider the set $\Sigma_x$ of all  
null directions outgoing from a given point $x\in\cal M$. This set is 
topologically a sphere ${\bf S}^2$ - the celestial sphere of an observer 
situated at $x$.
\footnote{We consider outgoing directions from $x$. In this sense directions 
generated by e.g. $k$ and $-k$ 
are considered to be different and two 
vector fields generate the same direction if an only if they differ by a 
multiple of a positive real function on $\cal M$.}  
The points 
of this sphere can be parametrized by a complex number $z$ belonging 
to the Argand plane ${\bf C}\cup\{\infty\}$. 
A direction associated with  $z\neq\infty$ is generated by a null vector 
\be 
k(z)= k +z\bar{z}l-z m -\bar{z}\bar{m}\label{eq:kz}. 
\ee 
With $z=\infty$ we associate a direction generated 
by the  vector $l$. Conversely, given a null direction outgoing from $x$ 
we find that it is either 
parallel to the vector $l$ or can be represented by only one null 
vector $k(z)$ such that $g(k(z),l)=-1$. It follows that such a vector $k(z)$ has 
necessarily the form (\ref{eq:kz}), and that it defines a certain $z\in\bf C$. If 
a direction is parallel to $l$ we associate with it $z=\infty$.\\
We define a fiber bundle 
${\cal P}=\displaystyle{\bigcup_{x\in\cal M}}\Sigma_x$ 
over $\cal M$, so  that the 2-dimensional spheres $\Sigma_x$ are its fibers. 
The canonical projection $\pi:{\cal P}\rightarrow{\cal M}$ is defined by  
$\pi(\Sigma_x)=x$. 
%I will call bundle $\cal P$ 
%as ``Penrose's twistor space'',  or `twistor bundle'. 
%This bundle posesses 
%quite a reach family of well defined geometrical objects, which can be 
%collected to the 
%so called `optical geometry' \cite{bi:ATOG}. Here we recall only those of 
%these 
The following geometrical objects existing on $\Pen$ are relevant in the 
present paper (see \cite{bi:optical} 
for details).
\begin{itemize}
\item[-] 
The Levi-Civita connection associated with the metric $g$ on $\M$ 
distinguishes a horizontal space in T$\cal P$. In this way for any 
point $p\in \cal P$ 
we have a natural splitting of its tangent space onto a direct sum 
T$_p{\cal P}=V_p\bigoplus H_p$, where $H_p$ is a 4-dimensional horizontal 
space and $V_p$ is a 2-dimensional vertical space. The vertical space 
$V_p$ is tangent to the fiber $\Sigma_{\pi(p)}$ at the point $p$. Thus $V_p$ 
has a natural complex structure related to the complex structure on the sphere 
${\bf S}^2$. The complexification of $V_p$ splits into eigenspaces $V^+_p$ and 
$V^-_p$ of this complex structure. We have a horizontal 
lift $\tilde{v}$ of any 
vector $v$ from $\pi (p)\in\M$ to $\Pen$. This is such a vector 
$\tilde{v}$ at $p$ that $\tilde{v}\in H_p$ and $\pi_* (\tilde{v})=v$.
\item[-] 
A Lorentzian metric $\tilde{g}$ can be defined on $\cal P$ 
by the requirements that
\begin{itemize}
\item[{\it a)}] the scalar product of any two horizontal vectors determined by 
$\tilde{g}$ is the same as the scalar product with respect to 
$g$ of their push 
forwards to 
$\cal M$,
\item[{\it b)}] the scalar product of any two vertical vectors with respect to $\tilde{g}$ 
is equal to their scalar product in the natural metric on the 2-dimensional 
sphere (this is consistent since vertical vectors can be considered 
tangent vectors to ${\bf S}^2$),
\item[{\it c)}] any two vectors such that one is horizontal and the other 
is vertical are orthogonal with respect to $\tilde{g}$.
\end{itemize}
\item[-] There is a natural congruence of oriented lines on $\Pen$ which is 
tangent to the horizontal lifts 
of null directions from $\cal M$. It is defined by the following 
recipe. Take any null vector $k$ at $x\in\cal M$. This represents 
a certain null direction $p(k)$ outgoing from $x$. Correspondingly, 
this defines a point $p=p(k)$ in the fiber $\pi^{-1}(x)$. Lift $k$ 
horizontally to $p$. This defines $\tilde{k}$ which generates a 
certain direction 
outgoing from $p\in\cal P$. Repeating this procedure for all 
directions outgoing from $x\in\cal M$ we attach to any point 
of $\pi^{-1}(x)$ a unique direction. If we do this for all points of 
$\cal M$, we define a field of directions on 
$\cal P$ which, according to its construction and the properties of 
$\tilde{g}$, is null. Since we considered outgoing null directions the 
integral curves of this field are oriented. They form the desired null 
congruence. This congruence is called the null spray on $\cal P$ 
\cite{bi:Spar}.
\end{itemize}

\noindent
Let $X$ 
be any nonvanishing vector field tangent to the null spray on 
$\cal P$. Let $\La'$ be a real 1-form on $\cal P$ defined by 
$\La' =-\tilde{g}(X)$. Since $X$ is defined up to a multiplication by 
a strictly positive real function on $\Pen$ then 
also $\Lambda'$ is specified up 
to a multiplication by a real strictly positive function, say  $1/A>0$, on 
$\cal P$ 
$$
\Lambda'\rightarrow\Lambda=\frac{1}{A}\Lambda', ~~~~A>0. 
$$
With the horizontal space in $\cal P$ 
one associates another 1-form. This is a complex 1-form $E'$ 
on $\cal P$ that for any $p\in\cal P$ we have $E'(H_p)=E'(V^-_p)=0$ and 
$E'\wedge\bar{E'}\neq 0$. This is defined up to a multiplication 
by a nonvanishing complex function, say $1/w$, on $\cal P$ 
$$
E'\rightarrow E=\frac{1}{w} E'.
$$ 
It is now easy to see that the metric $\tilde{g}$ on $\cal P$ can be 
expressed as 
$$\tilde{g}=2(\frac{1}{|w|^2}E'\bar{E'}+\Lambda' T' + F'\bar{F'})$$
for some choice of the 1-forms $T'$ (real) 
and $F'$ (complex) on $\cal P$. The above expression can be considered 
a definition of the forms $F'$ and $T'$. These are given up to the 
following 
transformations 
$$
F'\rightarrow F=\mbox{e}^{i\phi}(F'+\by\Lambda'),
$$
$$
T'\rightarrow T=A(T'+\by\bar{F}'+yF'+y\by\La'),
$$
where $\phi$ (real) and $y$ (complex) are some functions on $\Pen$.\\
It follows that the forms 
$(F', T',\La', E')$ may be expressed in terms of the ordered null cotetrad (\ref{eq:met}) 
and the corresponding connection by 
\beq
&\La'=L+z\bar{z}K+z\bar{M}+\bar{z}M,\nonumber\\
&F'=M+zK,\label{eq:formy}\\
&E'=\der z+\G_{32}+z(\G_{21}+\G_{43})+z^2\G_{14},\nonumber\\
&T'=K,\nonumber
\eeq
where $z$ is given by (\ref{eq:kz}) and we omit the pull back symbol $\pi^*$ 
in expressions such as 
$\pi^*(\G_{32})$, etc.\\

\noindent
In this way we see that the bundle $\cal P$ of 
null directions of any space-time $\cal M$ is equipped with the class of six 
1-forms $[(F', \bar{F}', T',\La', E',\bar{E}')]$ defined above. Following 
Elie 
Cartan this class of forms can be used to study all the invariant properties 
of the underlying Lorentzian geometry. Thus, we
consider a class of 1-forms $[( F', \bar{F}', T',\La', E',\bar{E}')]$ 
on 
$\Pen$ with the following properties.
\begin{itemize}
\item[({\it i.})]
$\La'$, $T'$ are real- and $F'$, $E'$ are complex-valued 1-forms on $\Pen$
\item[({\it ii.})]
$F'\dz\bar{F}'\dz T'\dz\La'\dz E'\dz\bar{E}'\neq 0$ at each point $p$ of 
$\Pen$
\item[({\it iii.})]
Two sets of forms 
$( F, \bar{F}, T,\La, E,\bar{E})$ and 
$( F', \bar{F}', T',\La', E',\bar{E}')$ are in the same class iff 
\be
\Lambda=\frac{1}{A}\Lambda',\label{eq:l}
\ee
\be
F={\rm e}^{i\varphi}( F' +\by \Lambda')\label{eq:f}
\ee
\be
T=A(T'+\by\bar{F}'+yF'+y\by\Lambda' )\label{eq:k}
\ee
\be
E=\frac{1}{w} E'.\label{eq:e}
\ee
Here $A>0$, $\varphi$ (real) $y$, $w\neq 0$ (complex) are 
arbitrary functions on $\Pen$. 
\item[({\it iv.})] 
A particular set of forms that belong to the considered class is given 
explicitly by (\ref{eq:formy}).
\end{itemize}

\noindent
Given the  representation (\ref{eq:formy}) of the forms we calculate 
their differentials. These read as follows.
\be
\der\La'=\bar{E}'\dz F'+ E'\dz \bar{F}'+
[\g-\bar{\om}_z]\La'\dz F'+[\bar{\g}-\om_{\bar{z}}]\La'\dz 
\bar{F}'+[\om+\bar{\om}]\La'\dz T'\label{eq:derl},
\ee
\be                                                   
\der F'=E'\dz T'+[-\g_{\bar{z}}+2\bar{\g}_z+
\bar{\om}_{z\bar{z}}]F'\dz\La'-\bar{\g}_{\bar{z}}\bar{F}'\dz\La'+
[\bar{\om}_{\bar{z}}+\bar{\g}]\La'\dz T'+
[\bar{\g}+\om_{\bar{z}}]\bar{F}'\dz F'+
[\om -\bar{\om}]F'\dz T',\label{eq:derf}
\ee
\beq
&\der T'=-\g_{z\bar{z}}F'\dz\La'-\bar{\g}_{z\bar{z}}\bar{F}'\dz\La'-
[\g_{\bar{z}}+\bar{\g}_z]\La'\dz T'+[\bar{\om}_z-\om_z-2\g]T'\dz F'+
\nonumber\\
&[\bar{\g}_z-\g_{\bar{z}}+\bar{\om}_{z\bar{z}}-\om_{z\bar{z}}]F'\dz \bar{F}'
+[\om_{\bar{z}}-\bar{\om}_{\bar{z}}-2\bar{\g}]T'\dz \bar{F}'\label{eq:derk},
\eeq
\beq
&\der E'=2\g E'\dz F'+2\g_{\bar{z}}\La'\dz E'-2\om_{\bar{z}}E'\dz\bar{F}'+
2\om E'\dz T'+\nonumber\\
&\Phi T'\dz F'+\Psi T'\dz\bar{F}'-\frac{1}{2}\Phi_{\bar{z}\bar{z}}
\La'\dz\bar{F}'+
[\frac{1}{12}\Psi_{zz}+\frac{1}{12}R]F'\dz \La'-\label{eq:dere}\\
&\frac{1}{4}\Psi_z[T'\dz\La'+F'\dz\bar{F}']-\frac{1}{2}
\Phi_{\bar{z}}[T'\dz\La'-F'\dz \bar{F}'],
\nonumber
\eeq
where we have used the following abreviations:
\be
2\g=\G_{211}+\G_{431}+2z\G_{141}-\bar{z}(\G_{214}+\G_{434})-2z\bar{z}
\G_{144},
\label{eq:gg}
\ee
\beq
&2\om=\G_{213}+\G_{433}+z(2\G_{143}-\G_{211}-\G_{431})-
\bar{z}(\G_{212}+\G_{432})+\\
&z\bar{z}(\G_{214}+\G_{434}-2\G_{142})-2z^2\G_{141}+2z^2\bar{z}\G_{144},
\nonumber
\eeq
\be
\Phi=\frac{1}{2}S_{33} - \bar{z}S_{23} -
zS_{13}  + z\bar{z} (S_{12}+S_{34}) +\frac{1}{2} \bar{z}^2 S_{22}
+\frac{1}{2} z^2 S_{11} - \bar{z}^2 zS_{24} - z^2\bar{z}S_{14} + 
\frac{1}{2}z^2\bar{z}^2S_{44}, 
\ee
\be
\Psi =\Psi_4-4\Psi_3z+6\Psi_2z^2-4\Psi_1z^3+
\Psi_0z^4,\label{eq:pp}
\ee
and the subscript $z$ (or $\bar{z}$) denotes the derivative with respect 
to $z$ (or $\bar{z}$).\\
It follows that the differential of $E'$ carries all the information about 
the Ricci tensor and the Weyl coefficients $\Psi_\mu$. In particular, the equation 
\be
\der E'\dz\La'\dz \bar{F}'\dz E'\equiv 0\label{eq:ei}, 
\ee
which is the same as
\be
\Phi\equiv 0,
\ee
is equivalent to the Einstein equations for the 4-metric. Similarly, 
\be
\der E'\dz\La'\dz F'\dz E'\equiv 0\label{eq:we}, 
\ee
which is the same as
\be
\Psi\equiv 0,
\ee
is equivalent to the conformal flatness of the metric. The equation 
\be
\Psi =0
\ee
is also important. It has at most four solutions at each fiber in $\Pen$ over 
a given point $x\in\cal M$. These four points, via (\ref{eq:kz}), 
correspond to four principal null directions at $x$.\\

\noindent
One easily discovers that both of the equations (\ref{eq:ei}) and 
(\ref{eq:we}) are 
invariant under the transformations (\ref{eq:l})-(\ref{eq:e}) of the forms. 
On the other hand, in the differentials of the six considered 1-forms 
there are terms which may be transformed to zero by an appropriate choice of 
the gauge (\ref{eq:l})-(\ref{eq:e}). Our aim now will be to use this gauge to 
get the simplest possible form of the differentials 
(\ref{eq:derl})-(\ref{eq:dere}).\\
We start our analysis with the forms $(F,\bar{F},T,\La,E,\bar{E})$ of  
(\ref{eq:l})-(\ref{eq:e}), in which $(F',\bar{F}',T',$ $\La',E',\bar{E}')$ 
are given 
by (\ref{eq:formy}).\\ 
%We hope that differentials of  $(\La',F',\bar{F}',K',E',\bar{E}')$ will 
%have nicer form than those of $(F,\bar{F},T,\La,E,\bar{E})$.\\
From the geometric point of view the forms 
$(F,\bar{F},T,\La,E,\bar{E})$ live on a manifold $\cal C'$ that has 
higher dimension than $\Pen$. Actually, if $p$ denotes a 
generic point of $\Pen$, then  
$\cal C'$ may be parametrized by 
$(p,A,\varphi,y,\bar{y},w,\bar{w})$. 
Thus $\cal C'$ is 12-dimensional and 
the forms $(F,\bar{F},T,\La,E,\bar{E})$ 
are well defined on it.\\
Calculating the differential of $\La$ on $\cal C'$ we find that
$$
\der\La=-\der{\rm log}A\dz\La+A^{-1}\{E'\dz\bar{F}'+\bar{E}'\dz F'
+[\g-\bar{\om}_z]\La'\dz F'+[\bar{\g}-\om_{\bar{z}}]\La'\dz \bar{F}'+
[\om+\bar{\om}]\La'\dz T'\}.
$$
This equation suggests the introduction of an auxiliary real-valued 1-form 
\be
\Omega+\bar{\Omega} =-\der{\rm log}A+s_1F'+\bar{s}_1\bar{F}'+s_3T'+
s_4\La'+s_5E'+\bar{s}_5\bar{E}'.
\ee
The functional coefficients $s_3,s_4$ (real) and $s_1,s_5$ (complex) are 
for the moment arbitrary. They may be used to eliminate some of the terms in 
the differential of $\La$. Indeed, using $\Omega+\bar{\Omega}$ we can 
rewrite the differential of $\La$ in the form
$$
\der\La=(\Omega+\bar{\Omega})\dz\La+
\frac{w{\rm e}^{i\varphi}}{A}E\dz\bar{F}+
\frac{\bar{w}{\rm e}^{-i\varphi}}{A}\bar{E}\dz F+
$$
$$
\La\dz\{[\g-\bar{\om}_z+s_1]F'+[\bar{\g}-\om_{\bar{z}}+\bar{s}_1]\bar{F}'+
[\om+\bar{\om}+s_3]T'+[y+s_5]E'+[\by+\bar{s}_5]\bar{E}'\}.
$$
This suggests the following choice of $s_1, s_2$ and $s_5$:
\be
s_1=-\g+\bar{\om}_z~~~s_3=-\om-\bar{\om},~~~s_5=-y.
\ee
With this choice the differential of $\La$ assumes the form 
$$
\der\La=(\Omega+\bar{\Omega})\dz\La+
\frac{w{\rm e}^{i\varphi}}{A}E\dz \bar{F}
+\frac{\bar{w}{\rm e}^{-i\varphi}}{A}\bar{E}\dz F.
$$
Now, we can make the first gauge fixing condition 
\be
w=A{\rm e}^{-i\varphi}.\label{eq:xi}
\ee
This brings the differential of $\La'$ to the simplest possible form 
\be
\der\La=(\Omega+\bar{\Omega})\dz\La+E\dz \bar{F}+\bar{E}\dz F.
\label{eq:l1}
\ee
Note that the choice 
(\ref{eq:xi}) uniquely subordinates $A$ and ${\rm e}^{i\varphi}$ to $w$. 
Explicitely, $A=|w|$, ${\rm e}^{i\varphi}=|w|/w$.   
Thus, after this choice, we have  
\be
\La=\frac{1}{|w|}\La'\label{eq:cccc}
\ee
\be
F=\frac{|w|}{w}(F'+\by\La')\label{eq:pip}
\ee
\be
T=|w|(T'+\by\bar{F}'+yF'+y\by\La')
\ee
\be
E=\frac{1}{w}E'.\label{eq:en}\label{eq:ccccc}
\ee
It is now clear that the set of all the  
forms $(F,\bar{F},T,\La,E,\bar{E})$ is well defined on the 
10-dimensional manifold $\cal C$ parametrized by 
$(p,w,\bar{w},y,\by)$. The price paid for the passage from 
$\cal C'$ to $\cal C$ is the introduction of a form 
\be
\Om+\bar{\Om}=-\der{\rm log}|w|+s_4\La'-(\om+\bar{\om})T'+
(\bar{\om}_z-\g)F'+
(\om_{\bar{z}}-\bar{\g})\bar{F}'-yE'-\by\bar{E}',
\ee
which is still not fully determined, since the real function $s_4$ is 
still arbitrary.\\

\noindent
We now pass to the analysis of $\der F$, with $F$ being given by 
(\ref{eq:pip}).\\ 
One easily calculates that
\be
\der F=\der{\rm log}\frac{|w|}{w}\dz F+\bar{w}\der\by\dz \La'+
\frac{|w|}{w}(\der F'+\by\der\La' ).
\ee
This suggests the introduction of two other auxiliary forms, $\Om-\bar{\Om}$ 
and $\G$, on $\cal C$, which are given by 
\be
\Om-\bar{\Om}=\der{\rm log}\frac{|w|}{w} +b_1F'-\bar{b}_1\bar{F}'+
b_3 T'+b_4\La'+b_5E'-\bar{b}_5\bar{E}',
\ee
\be
\G=w[\der y+c_1F'+c_2\bar{F}'+c_3 T'+c_4\La'+c_5E'+c_6\bar{E}'].
\ee
Here $b_3,b_4$ (purely imaginary) and $b_1,b_5,c_i$ ($i$=1,2,...6) 
(complex) are functions on $\cal C$ which should be determined.\\
Lenghty, but straightforward calculations lead to the following Lemma.
\bl~\\
If equation (\ref{eq:l1}) for ${\rm d}\La$ is satisfied then the conditions 
\be
{\rm d}F=(\Om -\bar{\Om})\dz F+\bar{\G}\dz\La+E\dz T\label{eq:f1}
\ee
\be
{\rm d}T=T\dz(\Om +\bar{\Om}) +\bar{\G}\dz\bar{F}+\G\dz F\label{eq:k1}
\ee
uniquely determine $s_4,~b_\mu,~c_i$ ($\mu$ =1,3,4,5; i=1,2,...,6) and, 
thus, the forms $\Om$ and $\G$. Explicitly, if $\La$, $F$, $T$, $E$ are 
given 
by (\ref{eq:cccc})-(\ref{eq:ccccc}) respectively, then   
\be
\Om =-\frac{1}{2}\frac{{\rm d}{w}}{w}+\g_{\bar{z}}\La'-
\g F'+\om_{\bar{z}}\bar{F}'-\om T'-yE'\label{eq:om}
\ee
\be
\G =w [{\rm d} y+(\g_{z\bar{z}}+2y\g_{\bz})\La'+
(\g_{\bz}+\om_{z\bar{z}}+2y\om_{\bz})\bar{F}'-
(\g_z+2y\g)F'-(\g+\om_z+2y\om)T'-y ^2E'].\label{eq:ga}
\ee
\el
Thus, on $\cal C$ the 1-forms 
$(F,\bar{F},T,\La,E,\bar{E},\Om,\bar{\Om},\G,\bar{\G})$ given by 
(\ref{eq:cccc})-(\ref{eq:ccccc}), (\ref{eq:formy}), 
(\ref{eq:om}), (\ref{eq:ga}) are well 
defined. They satisfy the differential equations  
(\ref{eq:l1}), (\ref{eq:f1})-(\ref{eq:k1}). From now on we analyze these 
forms.\\
Straightforward calculations lead to the following expression for the 
differential of $\der E$.
$$
\der E=2\Om\dz E+\frac{1}{|w|^2}\Phi T\dz F+
\frac{\bar{w}}{w}[\frac{1}{2}\Phi_{\bar{z}\bar{z}}+
\Phi_{\bar{z}}\by+\Phi\by^2]\bar{F}\dz \La+
$$
\be
\frac{1}{w}[\Psi y-\Phi\by+\frac{1}{4}\Psi_z-\frac{1}{2}
\Phi_{\bar{z}}]\bar{F}\dz F-\frac{1}{w}
[\Psi y+\Phi\by+\frac{1}{4}\Psi_z+\frac{1}{2}
\Phi_{\bar{z}}]T\dz \La+
\label{eq:e1}
\ee
$$
\frac{1}{w^2}\Psi T\dz\bar{F}+[\frac{1}{12}\Psi_{zz}+
\frac{1}{12}R+\frac{1}{2}\Psi_z y+\Psi y^2]F\dz \La.
$$
The following three cases are of particular interest.
\begin{itemize}
\item[({\bf A})] The metric $g$ of the 4-manifold $\cal M$ satisfies 
Einstein equations $R_{ij}=\la g_{ij}$ and is conformally non-flat. 
This case is characterized by $\Phi\equiv 0$ and $\Psi\not\equiv 0$.
\item[({\bf B})] The metric $g$ is conformally flat but not Einstein. 
This case corresponds to $\Psi\equiv 0$, $\Phi\not\equiv 0$.
\item[({\bf C})] The metric $g$ is of constant curvature. This means that 
$\Psi\equiv\Phi\equiv 0$.
\end{itemize}
Only in the first two cases is there a unique way of fixing the gauge 
for $(F,\bar{F},T,\La,$ $E,\bar{E})$. Thus in these two cases it is 
possible to reduce the system of 1-forms from $\cal C$ back to $\Pen$. Such 
a reduction corresponds to an appropriate 
choice of $y$ and $w$. As usual this choice will be such that it 
implies the vanishing of certain well defined terms in (\ref{eq:e1}). Such 
an approach is impossible in case ({\bf C}), since in this case there is an 
immediate reduction of (\ref{eq:e1}) to 
\be
\der E=2\Om\dz E+\frac{1}{12}R~F\dz\La.
\ee

\noindent
From now on we consider the case ({\bf A}) where the metric is not conformally flat and satisfies 
Einstein's equations. Imposing the restrictions ({\bf A}) on (\ref{eq:e1}) we immediately see 
that  
$$
\der E=2\Om\dz E-\frac{1}{w}
[\Psi y +\frac{1}{4}\Psi_z][T\dz \La+F\dz\bar{F}]+
$$
$$
-\frac{1}{w^2}\Psi T\dz\bar{F}+[\frac{1}{12}\Psi_{zz}+
\frac{1}{12}R+\frac{1}{2}\Psi_z y+\Psi y^2]F\dz\La.
$$
Assuming that $\Psi\neq 0$ and making the choice 
\be
y=-\frac{1}{4}\frac{\Psi_{z}}{\Psi}\label{eq:eta}
\ee
we bring $\der E$ to the form 
$$
\der E=2\Om\dz E-
\frac{1}{w^2}\Psi \bar{F}\dz T+[\frac{1}{12}\Psi_{zz}-\frac{1}{16}
\frac{\Psi_z^2}{\Psi}+\frac{1}{12}R]F\dz\La.
$$
Now the last gauge fixing condition can be made by demanding that 
\be
\w^2=-\Psi.\label{eq:2}
\ee
This determines $w$ up to a sign
\be
w=\pm i(\Psi)^{\frac{1}{2}}\label{eq:af}.
\ee 
Now, the expressions (\ref{eq:eta}) and (\ref{eq:af}) can be substituted into 
the 1-forms $(F,\bar{F},T,\La,E,$ 
$\bar{E},\Om,$ $\bar{\Om},\G,\bar{\G})$ 
given by (\ref{eq:cccc})-(\ref{eq:ccccc}), (\ref{eq:formy}), 
(\ref{eq:om}), (\ref{eq:ga}). After such a substitution the 
dependence of $y,\by,w,\bar{w}$ disappears from the forms. Thus 
they project to $\Pen$ where they are defined uniquely up to signs. This 
shows that in the case of the Einstein 4-metric we are able 
to fix the freedom in the choice of our initial 1-forms of 
(\ref{eq:l})-(\ref{eq:e}), everywhere on $\Pen$ except at points where $\Psi$ 
vanishes. As we know such vanishing occurs on sections of $\Pen$ 
corresponding to the principal null directions on $\cal M$. It is possible 
to overpass this difficulty by changing the topology of each fibre of 
$\Pen$. This possibility was studied by one of us in Ref. \cite{bi:einstein}.  Summing up we have the following theorem.
\bt~\\
Let $\cal M$ be a 4-dimensional Lorentzian manifold and let $\Pen $ be its 
corresponding bundle of null directions. Suppose that the metric $g$ on 
$\cal M$ satisfies the Einstein 
equations $R_{ij}=\la g_{ij}$ and is not conformally flat. 
Then on $\Pen$, apart the points 
that correspond to principal null directions, there exist 
prefered forms $(F,\bar{F},T,\La,E,\bar{E})$, which are in the 
class (\ref{eq:l})-(\ref{eq:e}), forms 
$(\Om,\bar{\Om},\G,\bar{\G})$ and a function $\al$ such that 
$$
{\rm d}\La=(\Omega+\bar{\Omega})\dz\La+E\dz\bar{F}+\bar{E}\dz F$$
\be
{\rm d}F=(\Om -\bar{\Om})\dz F+\bar{\G}\dz\La+E\dz T\label{eq:sys}
\ee
$$
{\rm d}T=T\dz(\Om +\bar{\Om}) +\bar{\G}\dz\bar{F}+\G\dz F
$$
$$
{\rm d}E=2\Om\dz E+\bar{F}\dz T+\al \La\dz F.
$$
The forms are given by 
$$
\La=\frac{1}{|\Psi|^{\frac{1}{2}}}\La',
$$
$$
F=\eps i(\frac{\bar{\Psi}}{\Psi})^{\frac{1}{4}}~[~F'-\frac{1}{4}
({\rm log}\bar{\Psi})_{\bar{z}}\La'~],~~~~~~~~~\eps=\pm 1,
$$ 
$$
T=|\Psi|^{\frac{1}{2}}[T'-\frac{1}{4}({\rm log}\Psi )_zF'
-\frac{1}{4}({\rm log}\bar{\Psi} )_{\bar{z}}\bar{F}'
+\frac{1}{16}|({\rm log}\Psi )_z|^2\La'],
$$
$$
E=\frac{\eps i}{|\Psi|^{\frac{1}{2}}}(\frac{\bar{\Psi}}{\Psi})^
{\frac{1}{4}}E',
$$
$$
\Om =-\frac{1}{4}{\rm dlog}\Psi+\g_{\bar{z}}\La'-
\g F'+\om_{\bar{z}}\bar{F}'-\om T'+\frac{1}{4}({\rm log}\Psi)_zE',
$$
$$
\G =-\eps i|\Psi|^\frac{1}{2}(\frac{\Psi}{\bar{\Psi}})^
\frac{1}{4}[-\frac{1}{4}{\rm d}({\rm log}\Psi)_z+
(\g_{z\bar{z}}-\frac{1}{2}({\rm log}\Psi)_z
\g_{\bz})\La'+(\g_{\bz}+\om_{z\bar{z}}-\frac{1}{2}({\rm log}\Psi)_z
\om_{\bz})\bar{F}'-
$$
$$
(\g_z-\frac{1}{2}({\rm log}\Psi)_z\g)
F'-(\g+\om_z-\frac{1}{2}({\rm log}\Psi)_z\om)T'-\frac{1}{16}({\rm log}
\Psi)_z^{~2}E'~],
$$
where $(\La',F',T',E')$ are given by (\ref{eq:formy}),  
$\g,\om,\Psi$ are those of (\ref{eq:gg})-(\ref{eq:pp}) and $\al$ is 
given by 
$$
\al=\frac{1}{16}\frac{\Psi_z^2}{\Psi}-\frac{1}{12}\Psi_{zz}-\frac{1}{12}R.
$$   
\et~\\
The forms 
$(F,\bar{F},T,\La,E,\bar{E},\Om,\bar{\Om},\G,\bar{\G})$ that appear 
in the above theorem will be called the Cartan invariant 1-forms for 
a Lorentzian conformally non-flat Einstein manifold. Together with the 
function $\al$, which we call the Cartan invariant function, they may be 
used to determine whether two given metrics are locally isometrically 
equivalent.\\

\section{Remarks on the equivalence problem}

\noindent 
Cartan's approach to the question of determining whether or not two 
given metrics are isometrically equivalent can be given a useful formulation 
in the context of the previous section. Here we outline the way in which 
Theorem 2 can be used to do this.\\

\noindent
Suppose that we are given two Lorentzian metrics $g$ and $\hat{g}$ 
on two 4-manifolds $\cal M$ and $\hat{\cal M}$. The metrics are assumed to 
be Einstein and conformally non-flat. Suppose now that there exists a local 
isometry between $g$ and $\hat{g}$, that is a local diffeomorphism 
$\phi:{\cal M}\to \hat{\cal M}$ such that $\phi^*\hat{g}=g$. 
Taking an ordered null cotetrad 
$(\hat{M},\bar{\hat{M}},\hat{K},\hat{L})$ on $\hat{\cal M}$ 
and applying to it $\phi^*$  we get the 1-forms  
\be 
M=\phi^*(\hat{M}),~~~~\bar{M}=\phi^*(\bar{\hat{M}}),~~~~
K=\phi^*(\hat{K}),~~~~L=\phi^*(\hat{L}).\label{eq:867}
\ee 
Due to the isometric property of $\phi$ we find that $(M,\bar{M},K,L)$ 
constitutes a null cotetrad for $g$ on $\cal M$. Now we use the cotetrads 
$(M,\bar{M},K,L)$ on $\cal M$ and $(\hat{M},\bar{\hat{M}},\hat{K},\hat{L})$
on $\hat{\cal M}$ to calculate the Cartan invariants on the corresponding 
bundles of null directions $\Pen$ and $\hat{\Pen}$. Let $z$ and 
$\hat{z}$ be fiber coordinates on $\Pen$ and $\hat{\Pen}$ 
related to the cotetrads  
$(M,\bar{M},K,L)$ and $(\hat{M},\bar{\hat{M}},\hat{K},\hat{L})$
by the formulae analogous to (\ref{eq:kz}). Then Theorem 2 yields two 
sets of the Cartan invariant 1-forms: 
$(F,\bar{F},T,\La,E,\bar{E},\Om,\bar{\Om},\G,\bar{\G})$ on $\Pen$ 
and  $(\hat{F},\hat{\bar{F}},\hat{T},\hat{\La},\hat{E},\hat{\bar{E}},
\hat{\Om},\hat{\bar{\Om}},\hat{\G},\hat{\bar{\G}})$ on $\hat{\Pen}$. 
In addition a pair of 
Cartan invariants $\al$ and $\hat{\al}$ may be easily calculated. It follows 
from the definition of the Cartan invariants and  
(\ref{eq:867}) that  the map $\hat{p}:\Pen\to\hat{\Pen}$ defined by 
$\hat{p}(x^i,z,\bar{z})=(\phi(x^i),z,\bar{z})$ has the property 
that 
\be
\hat{p}^*(\hat{\La})=\La,~~~~ \hat{p}^*(\hat{T})=T,~~~~
\hat{p}^*(\hat{F})=\pm F,~~~~ \hat{p}^*(\hat{E})=\pm E,
\label{eq:ci1}
\ee
\be 
\hat{p}^*(\hat{\Om})=\Om~~~~ \hat{p}^*(\hat{\G})=\pm\G,
\label{eq:ci2}
\ee
\be
\hat{p}^*(\hat{\al})=\al.
\label{eq:ci3}
\ee
This proves the following Proposition.    
\bp~\\
Any (local) diffeomorphism 
$\phi:{\cal M}\to \hat{\cal M}$ which is an isometry between $g$ and 
$\hat{g}$ 
generates a (local) diffeomorphism 
$\hat{p}:\Pen\to\hat{\Pen}$ which satisfies (\ref{eq:ci1})-(\ref{eq:ci3}).
\ep
To prove the converse we need the following Lemma.
\bl~\\
A diffeomorphism $\hat{p}:\Pen\to\hat{\Pen}$ satisfying 
(\ref{eq:ci1})-(\ref{eq:ci3}) induces a diffeomorphism 
$\phi:{\cal M}\to\hat{\cal M}$ such that the following diagram\\
\be
\begin{CD}
\Pen @>\hat{p}>>\hat{\Pen}\\
@VV\pi V @VV\hat{\pi}V \\
{\cal M}@>\phi >> \hat{\cal M}\\
\end{CD}
\label{eq:diag}
\ee\\
commutes.
\el
Proof.\\
$\Pen$ is foliated by the fibers $\Sigma_x$ of the fibration 
$\pi:\Pen\to\cal M$. This foliation, which we denote by $\cal V$, 
is such that each of the forms $(F,\bar{F},T,\La)$ vanishes 
when restricted to its leaves $\Sigma_x$ \footnote{This is consistent, 
since the forms $(F,\bar{F},T,\La)$ constitute a closed differential 
ideal due to the equations (\ref{eq:sys}).}. An analogous foliation 
$\hat{\cal V}$ exists on $\hat{\Pen}$.  
Our aim is to prove that two points from the same leaf of $\cal V$ can not 
be transformed by $\hat{p}$ to different leaves of $\hat{\cal V}$. 
To do this, observe that the diffeomorphism property of $\hat{p}$ 
implies that the leaves of the foliation $\cal V$ are transformed by 
$\hat{p}$ to non-intersecting 2-dimensional submanifolds of $\hat{\Pen}$. 
Moreover, equations (\ref{eq:ci1}) guarantee that each of the forms 
$(\hat{F},\bar{\hat{F}},\hat{T},\hat{\La})$ 
identically vanishes when restricted to any of these 2-manifolds. Thus, 
the set of all $\hat{p}(\Sigma_x)$, $x\in\cal M$,    
defines a foliation $\hat{p}({\cal V})$ of $\hat{\Pen}$ which possesses all 
the properties of $\hat{\cal V}$. Due to the uniqueness of such foliation, 
which follows from the Frobenius theorem, 
$\hat{p}({\cal V})\equiv \hat{\cal V}$. This in particular means that 
points from a given leaf of $\cal V$ are transformed by $\hat{p}$ to the 
same leaf of $\hat{\cal V}$. Thus the map $\hat{p}:\Pen\to\hat{\Pen}$ 
projects to the map $\phi:{\cal M}\to\hat{\cal M}$. This map, by definition, 
has the property (\ref{eq:diag}). This proves the Lemma.\\

\noindent
Now, conversely to Proposition 3, we have 

\noindent
\bp~\\
If there exists a diffeomorphism 
$\hat{p}:{\cal P}\to \hat{\cal P}$, which satisfies equations 
(\ref{eq:ci1})-(\ref{eq:ci3}), then the metrics $g$ and $\hat{g}$ are 
isometrically equivalent. 
\ep
To prove this, consider $\phi$ of Lemma 2 and observe that 
diagram (\ref{eq:diag}) implies that 
\be
\hat{p}^*\hat{\pi}^*\hat{g}=\pi^*\phi^*\hat{g}.\label{eq:12}
\ee
On the other hand since 
$\hat{\pi}^*\hat{g}=2(\hat{F}\bar{\hat{F}}-\hat{T}\hat{\La})$ 
and $\pi^*g=2(F\bar{F}-T\La)$, 
applying (\ref{eq:ci1}) gives $\hat{p}^*\hat{\pi}^*\hat{g}=\pi^*g$. 
Comparison of this with (\ref{eq:12}) yields $\phi^*\hat{g}=g.$ 
This proves Proposition 4.\\

\noindent
It follows that we have the following algorithm for 
checking the local isometric equivalence of 4-metrics. 
\begin {itemize}
\item[{\bf (a)}] Calculate the Petrov types of the 
metrics $g$ and 
$\hat{g}$. 
If the Petrov types are different then the metrics are not equivalent. 
\item[{\bf (b)}] If the Petrov types are the same calculate the Cartan 
invariant 1-forms  
$(F,\bar{F},$ 
$T,\La, E,\bar{E},\Om,\bar{\Om},\G,\bar{\G})$ on $\Pen$ 
and  $(\hat{F},\hat{\bar{F}},\hat{T},\hat{\La}, \hat{E},\hat{\bar{E}},
\hat{\Om},\hat{\bar{\Om}},\hat{\G},\hat{\bar{\G}})$ on $\hat{\Pen}$. Also 
calculate the Cartan invariant functions $\al$ and $\hat{\al}$. 
\item[{\bf (c)}] Search for a diffeomorphism 
$\hat{p}:{\cal P}\to \hat{\cal P}$ 
which satisfies (\ref{eq:ci1})-(\ref{eq:ci3}). The metrics are 
(locally) equivalent if and only if such a $\hat{p}$ exists. 
\end{itemize}
To perform step {\bf (c)} one needs to 
solve differential equations such as, for example, $\hat{p}^*(\hat{T})=T$ 
for $\hat{p}$. This may be not easy. To avoid this difficulty the 
following alternative procedure can be used. Recall that the forms 
$(F,\bar{F},T,\La,E,\bar{E})$ (respectively, 
$(\hat{F},\hat{\bar{F}},\hat{T},\hat{\La},\hat{E},\hat{\bar{E}})$) 
are linearly independent at each point of $\Pen$ 
(respectively, $\hat{\Pen}$). We can therefore use the basis 
$(F,\bar{F},T,\La, E,\bar{E})$ (respectively, 
$(\hat{F},\hat{\bar{F}},\hat{T},\hat{\La},\hat{E},\hat{\bar{E}})$) 
to decompose the forms 
$(\Om,\bar{\Om},\G,\bar{\G})$  
(respectively, $(\hat{\Om},\hat{\bar{\Om}},\hat{\G},\hat{\bar{\G}})$) onto 
them. 
These decompositions  
$$
\Om=\om_1F+\om_2\bar{F}+\om_3T+\om_4\La+\om_5E+\om_6\bar{E}
$$
$$
\G=\g_1F+\g_2\bar{F}+\g_3T+\g_4\La+\g_5E+\g_6\bar{E}
$$
define coefficients $\om_i,~\g_i$, and the analogous coefficients 
$\hat{\om}_i,~\hat{\g}_i$ for 
$\hat{\Om},~\hat{\G}$.
%Analogous formulae for $\hat{\Om}$ and $\hat{\G}$ define ${\hat{\om}}_i$ and 
%${\hat{\g}}_i$. 
The functions $\om_i,~\g_i$, $i=1,2,...6$ will be called the higher order 
Cartan invariant functions for the Lorentzian conformally non-flat 
Einstein metric $g$.\\
\noindent
It is easy to see that some of the higher order Cartan 
invariant functions 
identically vanish. Indeed, from the definitions of $\Om$ and $\G$ 
given in Theorem 2 one easily finds that 
$$
\om_5\equiv\om_6\equiv\g_6\equiv 0.
$$
It is also straightforward to see that 
$$
\g_5+3\al+\frac{1}{4}R\equiv 0.
$$
By using the Bianchi identities for $\Psi$ one also obtains the equations 
$$
4\om_4+\g_2\equiv 4\om_1+\g_3\equiv 0.
$$
%We do not present a proof of this identity here 
%since an altenative proof of this fact will be given in Section 6.\\

\noindent
Hence we can always write $\Om$ and $\G$ in the form 
\beq
&\Om=\om_1F+\om_2\bar{F}+\om_3T+\om_4\La,\nonumber\\
&\G=\g_1F-4\om_4\bar{F}-4\om_1T+\g_4\La-(3\al+\frac{1}{4}R)E,\nonumber
\eeq
(compare with (\ref{eq:domy2})).\\
Thus the relevant Cartan invariant functions are: $\al$, 
$\om_i$, $i=1,2,3,4$, $\g_1$, $\g_3$ and their complex conjugates. $R$ is 
a constant invariant. In terms of these invariants 
the conditions (\ref{eq:ci2})-(\ref{eq:ci3}) for $\hat{p}$ may be rewriten  
in the form  
%(\ref{eq:ci1}) and to 
\beq
&\hat{p}^*(\hat{\al})=\al,\nonumber\\
&\hat{p}^*(\hat{\om}_1)=\pm\om_1,~~~
\hat{p}*(\hat{\om}_2)=\pm\om_2,~~~\hat{p}^*(\hat{\om}_3)=\om_3,~~~
\hat{p}^*(\hat{\om}_4)=\om_4,~~~
\label{eq:ci4}\\
&\hat{p}^*(\hat{\g}_1)=\g_1,~~~\hat{p}^*(\hat{\g}_4)=\pm\g_4\nonumber\\
&R=\hat{R}.\nonumber
\eeq
It follows that (as is well known) metrics with different scalar 
curvatures $R$ and $\hat{R}$ are always nonisometric. If $R=\hat{R}$ then 
equations (\ref{eq:ci1}) and  (\ref{eq:ci4}) are equivalent to the system 
(\ref{eq:ci1})-(\ref{eq:ci3}). However, the system 
(\ref{eq:ci1})-(\ref{eq:ci3}) includes only two non-differential 
equations (\ref{eq:ci3}) and $\overline{(\ref{eq:ci3})}$ for 
$\hat{p}$. On the other hand the system (\ref{eq:ci1}), (\ref{eq:ci4}) 
includes 14 non-differential equations. These are precisely (\ref{eq:ci4}) 
and their complex conjugates. \\
%\beq
%&\hat{\al}(\hat{p}^i)=\al(p^i),\nonumber\\
%&\hat{\om_1}(\hat{p}^i)=\om_1(p^i),~~~
%\hat{\om_2}(\hat{p}^i)=\pm\om_2(p^i),~~~
%\hat{\om_3}(\hat{p}^i)=\pm\om_3(p^i),\nonumber\\
%&\hat{\om_4}(\hat{p}^i)=\om_4(p^i),~~~
%\hat{\om_5}(\hat{p}^i)=\pm\om_5(p^i),~~~
%\hat{\om_6}(\hat{p}^i)=\pm\om_6(p^i),\label{eq:ci4}\\
%&\hat{\g_1}(\hat{p}^i)=\pm\g_1(p^i),~~~
%\hat{\g_2}(\hat{p}^i)=\g_2(p^i),~~~
%\hat{\g_3}(\hat{p}^i)=\g_3(p^i),\nonumber\\
%&\hat{\g_4}(\hat{p}^i)=\pm\g_4(p^i),~~~
%\hat{\g_5}(\hat{p}^i)=\g_5(p^i),~~~\hat{\g_6}(\hat{p}^i)=\g_6(p^i),\nonumber
%\eeq
%and their complex conjugos. We see that to each Cartan invariant function 
%corresponds one equation from the above list.\\ 

\noindent
Now, suppose that 6 independent real functions, 
say $\hat{f}_1,\hat{f}_2,\hat{f}_3,\hat{f}_4,\hat{f}_5,\hat{f}_6$ 
($\der\hat{f}_1\dz\der \hat{f}_2\dz\der \hat{f}_3\dz\der \hat{f}_4\dz\der 
\hat{f}_5\dz\der \hat{f}_6\neq 0$), of the real and imaginary parts of  
the Cartan invariants 
$\hat{\al}$, ${\hat{\om}}_1$, ${\hat{\om}}_2$, ${\hat{\om}}_3$, 
${\hat{\om}}_4$, $\hat{\g}_1$, $\hat{\g}_4$ can be 
chosen near a point $\hat{p}_o\in\hat{\Pen}$. 
\footnote{In such cases it 
is convenient to use them as a coordinate system on $\hat{\Pen}$.} Taking 
the corresponding functions $f_1,f_2,f_3,f_4,f_5,f_6$ on $\Pen$ and using 
equations (\ref{eq:ci4}) we find that the map $\hat{p}$ must satisfy 
the six independent non-differential equations 
\be
\hat{p}^*\hat{f}_i=\eps f_i,~~~~i=1,2...,6.\label{eq:cii} 
\ee
Here $\eps$ may be either 1 or $- 1$ depending on which of 
the Cartan invariants we have used. It follows from the 
implicit function theorem that the six equations (\ref{eq:cii}) 
uniquely determine the desired map $\hat{p}^i=\hat{p}^i(p^j)$. Thus, in this 
case, to solve the equivalence problem for 
the two metrics we have to check whether $\hat{p}$ so determined
satisfies all the remaining equations (\ref{eq:ci4}) and   
the differential equations (\ref{eq:ci1}). If it does then 
the two metrics are isometrically equivalent, otherwise they are 
not. This solves the equivalence problem for the Lorentzian 
conformally non-flat Einstein metrics in the generic case. The discussion 
does not apply to the case when the number of independent functions 
among the Cartan invariants is less then six. We call such cases 
degenerate. These are more subtle and will be presented 
elsewhere.\footnote{Here we only note that if the number of 
independent Cartan   
invariant functions is less than six two cases may occur. Either all the 
Cartan invariant functions are constant or there exists at least one which 
is 
not constant. The former case will be totally analyzed in Section 8. 
In the latter case one takes the differential of the nonconstant 
Cartan invariant and decomposes it onto a basis of the Cartan invariant 
forms 
$(\hat{F},\hat{\bar{F}},\hat{T},\hat{\La}, \hat{E},\hat{\bar{E}})$. 
This produces new Cartan invariant functions of the next order. In generic 
cases one can use these to get new algebraic equations for $\hat{p}$. If in 
this way we are able to produce six independent algebraic equations for 
$\hat{p}$ then we return to the already discussed case. If not the 
procedure can be applied once more. There will be, of course, cases in 
which it is not possible to construct six independent algebraic equations 
for $\hat{p}$. This may occur if, for example, all the Cartan invariants 
depend only on one variable. In such cases there are symmetries and they 
may be analyzed by using group theoretical methods.}
%We 
%believe that if at the certain order it is not the case then all such 
%metrics can be classified. Thus, for the equivalence problem the worst case 
%would be, when at each step of the    
%procedure a new differentiation brought always only one Cartan invariant.  
%This gives a heuristic argument that to completely solve an equivalence 
%problem for two Lorentzian Einstein 4-metrics one 
%needs to differentiate the initial Cartan invariant at most 5 times. Since 
%this is of at most 3rd order in terms of the metric coefficients, one  
%gets the 
%of the preceding 
%section. This is the most interesting Einstein case.\\ 
%The cases ({\bf B}) and ({\bf C}) 
%are left to the reader.\\
% upper bond for the number $n$ of differentiations of the Einstein 
%4-metrics to get the set of invariants which allows for the algebraic 
%solution for the Cartan equivalence problem. This bond is $n\leq 8$.\\

\noindent
\section{Elliptic fibrations}
Suppose now that we have six 
1-forms $(F,\bar{F},T,\La, E,\bar{E})$ defined on an open set $\Pen_0$ 
of ${\bf R}^6$ which satisfy the differential system 
$({\cal I},\Pen_0)$ of Theorem 1. According to this theorem $\Pen_0$ is 
foliated by 2-dimensional leaves in such a way that it can be 
considered a fibration over the Einstein conformally non-flat space-time 
$\cal M$. Theorem 1 says nothing about the topology of the fibres of 
$\Pen_0$ since it deals with local solutions to differential equations 
(\ref{eq:sys1}). Thus, given a solution $(F,\bar{F},T,\La, E,\bar{E})$ 
of the system ${\cal I}$ on $\Pen_0$ we know only that $\Pen_0$ is foliated 
by leaves that in the generic case have the topology of an open disk in 
${\bf R}^2$. The question arises whether we can extend the solution 
$(F,\bar{F},T,\La, E,\bar{E})$ to a larger fibration $\tilde{\cal P}$ over 
$\cal M$ in such a way that its fibres contain fibres of $\Pen_0$, and have 
a more interesting topology than that of an open disks. It follows that given a 
solution there may be several such extensions. In this 
section we describe the most natural one, making more explicit the 
considerations of Ref.\cite{bi:einstein}. The other possibility is discussed 
in Section 7.\\

\noindent
A natural way of extending the fibres of $\Pen_0$ of Theorem 1 is as 
follows. Given $\Pen_0$ with the system $\cal I$ on it, one passes to the 
space-time $\cal M$ that is associated with it via Theorem 1. Then, using 
Theorem 2, one considers the bundle of null directions ${\cal P}$ for 
$\cal M$ and defines the Cartan invariant 1-forms 
$(F,\bar{F},T,\La, E,\bar{E})$ on it. These forms satisfy 
again the system of equations (\ref{eq:sys1}). Thus, we have an extension map 
$\phi$ from $\Pen_0$ with its fibres (say, of open disk topology) to 
${\cal P}$ with fibres being spheres of null directions. The only problem is 
that some of the forms $(F,\bar{F},T,\La, E,\bar{E})$ on ${\cal P}$ are 
defined only up to a sign (see Theorem 2). To avoid this double-valuedness of 
$F$ and $E$ we again need to extend the fibres of ${\cal P}$. This is done 
as follows.\\

\noindent      
Recall that the Cartan invariant 1-forms of the Theorem 2 are defined on 
$\Pen$ by the gauge fixing conditions (\ref{eq:eta}), (\ref{eq:2}). The 
first of these conditions makes no sense if  
$$\Psi =\Psi_4-4\Psi_3z+6\Psi_2z^2-4\Psi_1z^3+\Psi_0z^4$$
is zero. Here $z\in{\bf C}\cup\{\infty\}$ is a coordinate on a given fibre of 
$\Pen$. Thus in each fibre of $\Pen$ there are at most four points 
(which via 
(\ref{eq:kz}) correspond to principal null directions at the space-time point)
at which the above expression vanishes. Consider now the function 
$w=\sqrt{-\Psi}$ defined by condition (\ref{eq:2}). We analyze how $w$ 
changes when we pass along a small loop around a zero of $\Psi$ in a fibre.\\
%Moreover, some of these forms are doubly-valued due to the  
%appearence of $\eps$ in their definitions. R. Penrose 
%\cite{bi:Pen} suggests to change the topology of $\Pen$ in such a way 
%that all the forms will be single-valued and well defined everywhere 
%on such a new manifold. To deal with this sugestion we need some 
%preparations.\\ 
\noindent
We write $\Psi=c(z-z_1)(z-z_2)(z-z_3)(z-z_4)$, where the roots 
$z_i,~i=1,2,3,4$ are, in general, different. As is well known, if some of the 
roots $z_i$ coincide then the 4-metric is algebraically special. The 
resulting Petrov types of the 4-metric are:\\
\centerline{$z_1$, $z_2$, $z_3$, $z_4$ all different $\lr$ Petrov type I}
\centerline{$z_1=z_2$, $z_3, z_4$ different $\lr$ Petrov type II}
\centerline{$z_1=z_2\neq z_3=z_4$ different $\lr$ Petrov type D}
\centerline{$z_1=z_2=z_3$, $z_4$ different $\lr$ Petrov type III}
\centerline{$z_1=z_2=z_3=z_4$ $\lr$ Petrov type N}
Suppose now that we consider a space-time point for which the metric is of 
Petrov type I. Consider a loop 
\be
z(t)=z_i+\rho{\rm e}^{it},\label{eq:loop} 
\ee
$t\in [0,2\pi]$ around one of the roots $z_i$ (say $z_1$) of $\Psi$. The 
loop is supposed to be small ($\rho <<1$), so that the value of 
$\Psi =\Psi(t)$ at the points of this loop may be approximated by 
$\Psi=c(z_1-z_2)(z_1-z_3)(z_1-z_4)\rho{\rm e}^{it}$. 
We can write this fact as $\Psi=-\sigma_o{\rm e}^{i(t+t_o)}$, 
where $\sigma_o>0$ and $t_o$ are real constants. Substituting 
such a $\Psi$ in $w=\sqrt{-\Psi}$ we find that 
$w=\sqrt{\sigma_0}{\rm e}^{i\frac{t_0}{2}}{\rm e}^{i\frac{t}{2}}$. 
Suppose now that we change $t$ from 0 to $2\pi$. This 
corresponds to the passage from a point $z(0)$ to itself along the  
loop. Since $t/2$ changes its value only by $\pi$, then $w$ changes sign. 
To return to the initial value of $w$ we need to change $t$ from 0 to $4\pi$. 
This shows that the point $z_1$ is a double branch point for the function 
$w$. If the metric of the space-time point under consideration 
is algebraically general the same is true for $z_2,z_3,z_4$. To make the 
function $w$ single-valued we need to extend 
the $z$-space (a sphere) to a torus. This is obtained by considering two 
copies of the $z$-spaces (two spheres - 
the domains on which $w$ has plus and minus sign, respectively) each having 
two cuts between its points $z_1$-$z_2$ and $z_3$-$z_4$, say, with an 
appropriate identification of the cut edges.\\
The situation is a bit different if some of the roots $z_i$ of $\Psi$ are 
multiple. It is easy to see that after a passage along a loop around a double 
or quadruple root the function $w$ does not change sign, and that after a 
passage around a triple root $w$ changes a sign and returns to its original 
value only after the second turn. Considering two copies of spheres with 
apropriate cuts and identifications for all the possible situations we arrive 
to the following conclusion.\\
Let $x$ be a space-time point and ${\cal T}_x$ its space on which the function 
$w$ is single-valued. If the metric at $x$ is:
\begin{itemize}
\item[-] algebraically general, then ${\cal T}_x$ has topology of a 2-torus. 
\item[-] of Petrov type II, then ${\cal T}_x$ has topology of a 2-torus with one vanishing cycle.
\item[-] of Petrov type D, then ${\cal T}_x$ has topology of two 2-spheres 
touching each other in two different points.
\item[-] of Petrov type III, then ${\cal T}_x$ has topology of a 2-sphere with 
one singular point. 
\item[-] of Petrov type N, then ${\cal T}_x$ has topology of two 2-spheres 
touching each other in one point.
\end{itemize}

\noindent
Recall that the origin of the double-valuedness of the Cartan invariant 
forms was the double-valuedness of function $w$ which we used in their 
definition via (\ref{eq:2}). Thus it is clear that on a fibration $\pn$ 
whose fibre over a space-time point $x$ is ${\cal T}_x$ the 1-forms of 
Theorem 2 are 
single valued, and can have only simple singularities in at most four points 
at the fibre. This shows that the fibration $\Pen_0$ on which the system $\cal I$ 
of Theorem 1 is satisfied can be naturally extended to the fibration $\pn$ 
whose fibres over a space-time point $x$ are tori or their degenerate 
counterparts, depending on the algebraic type of the metric at $x$. Since 
fibres with the topology of tori are defined by an algebraic equation 
$w^2=\Psi_4-4\Psi_3z+6\Psi_2z^2-4\Psi_1z^3+\Psi_0z^4$, which can be identified 
with an elliptic curve (possibly degenerate) in ${\bf C}^2$, we call $\Pen$ an 
elliptic fibration \cite{bi:einstein}. It is clear that it constitutes a 
double branched cover of the bundle of null directions over the space-time.

\section{Generalized bundles of null directions and their symmetries}

\noindent
It is now clear that given the system $(\Pen_0,{\cal I})$ of Theorem 1 one has 
not enough information to determine the topology of the fibres of $\Pen_0$. We 
know that locally $\Pen_0$ has all the properties of the bundle of null 
directions over the associated Einstein space-time and that it can be further  
extended to the elliptic fibration of the preceding section. However other 
extensions are possible. In this section we consider examples of systems 
$(\Pen_0,{\cal I})$ of Theorem 1 that can have fibres with 
topologies different 
from those discussed so far. They correspond to the known Einstein spaces with 
6-dimensional groups of symmetries.\\

\noindent
Because of local equivalence of $\Pen_0$ and $\Pen$ we 
introduce the following definition.
 
\bd~\\
A real 6-dimensional 
manifold $\Pen$ equipped with forms $(F,\bar{F},T,\La, E,\bar{E})$ 
which satisfy the differential system (\ref{eq:sys1}) will be called a 
generalized Einstein bundle of null directions. 
\ed
In this section the letter $\Pen$ always denotes such bundles.\\ 

\noindent
Via Theorems 1 and 2 the forms 
$(F,\bar{F},T,\La, E,\bar{E})$ on $\Pen$, as well as the derived 
quantities $\al$, $\Om$ and $\G$, may be identified with the 
Cartan invariants for the associated Einstein space-time. Thus we will 
also call them the Cartan invariants for $\Pen$.\\

\bd~\\
We say that $\Pen$ is (locally) symmetric iff there exists a 
(local) diffeomorphism 
$\phi:\Pen\to \Pen$ which preserves the 1-forms 
$(F,\bar{F},T,\La,E,\bar{E})$, i.e. iff 
\be
\phi^*(F)=\pm F,~~~~\phi^*(T)=T,~~~~
\phi^*(\La)=\La,~~~~ \phi^*(E)=\pm E.
\label{eq:sym1}
\ee
A real vector field $\tilde{X}$ on $\Pen$ is a symmetry iff 
\be
{\cal L}_{\tilde{X}}F=0,~~~~{\cal L}_{\tilde{X}}T=0,~~~~
{\cal L}_{\tilde{X}}\La=0,~~~~{\cal L}_{\tilde{X}}E=0.
\label{eq:sym2}
\ee
\ed
The following Lemma shows that all of the Cartan invariants are preserved 
by a symmetry.
\bl
If $\tilde{X}$ is a symmetry of $\Pen$ then
\be
\tilde{X}(\al)=0,~~~{\cal L}_{\tilde{X}}\Om=0,~~~{\cal L}_{\tilde{X}}\G=0, 
\label{eq:sym3}
\ee
where $\Om$, $\G$ and $\al$ are the Cartan invariants of equations 
(\ref{eq:sys1}).
\el
To prove this we observe that (\ref{eq:sym2}) implies  
${\cal L}_{\tilde{X}}\der\La=0$, ${\cal L}_{\tilde{X}}\der F=0$, 
${\cal L}_{\tilde{X}}\der T=0$, ${\cal L}_{\tilde{X}}\der E=0$. Now, 
combining these equations with (\ref{eq:sys1}) we find that 
$$
[{\cal L}_{\tilde{X}}(\Om+\bar{\Om})]\dz \La=0,
$$
$$
[{\cal L}_{\tilde{X}}(\Om-\bar{\Om})]\dz F+[{\cal L}_{\tilde{X}}\bar{\G}]\dz\La=0,
$$
$$
-{\cal L}_{\tilde{X}}(\Om+\bar{\Om})\dz T+{\cal L}_{\tilde{X}}\bar{\G}\dz\bar{F}+
{\cal L}_{\tilde{X}}\G\dz F=0,
$$
$$
2[{\cal L}_{\tilde{X}}\Om]\dz E+\tilde{X}(\al)\La\dz F=0.
$$
Due to the independence of $(F,\bar{F},T,\La,E,\bar{E})$ the above 
equations imply (\ref{eq:sym3}). This finishes the proof of Lemma 3.\\ 

\noindent
\bd~\\
A symmetry is called vertical if it has the form 
$\tilde{X}=se+\bar{s}\bar{e}$, where $s$ is any complex-valued 
function on $\Pen$, and 
$(f,\bar{f},t,l,e,\bar{e})$ constitute a basis of vector fields on $\Pen$ 
dual to $(F,\bar{F},T,\La,E,\bar{E})$. 
\ed
\bl~\\
The only vertical symmetry of $\Pen$ is $\tilde{X}=0$.
\el
Proof.\\
Let $\tilde{X}$ be a symmetry of $\Pen$. Its general form is   
\be
\tilde{X}=x_1f+\bar{x}_1\bar{f}+x_3t+x_4l+x_5e+\bar{x}_5\bar{e},
\label{eq:ssym}
\ee
where $x_3,x_4$ are real functions. Using the symmetry conditions 
(\ref{eq:sym2}) and the differentials (\ref{eq:sys1}), 
(\ref{eq:domy1})-(\ref{eq:domy2}) we find that 
\be
\der x_1=(\bar{x}_7-x_7)F-x_5T-x_8\La+x_3E+x_1(\Om-\bar{\Om})+x_4\bar{\G},  
\label{eq:x1}
\ee
\be
\der x_3=-\bar{x}_8F-x_8\bar{F}+(x_7+\bar{x}_7)T-x_3(\Om+\bar{\Om})+\bar{x}_1
\bar{\G}+x_1\G,
\ee
\be
\der x_4=-\bar{x}_5F-x_5\bar{F}-(x_7+\bar{x}_7)\La+\bar{x}_1E+x_1\bar{E}+x_4
(\Om+\bar{\Om}),
\ee
\be
\der x_5=-\al x_4F+x_3\bar{F}-\bar{x}_1T+\al x_1\La-2 x_7E+2x_5\Om,
\label{eq:x5}
\ee
where $x_7$ and $x_8$ are functions whose form is not relevant here. 
Now, if the symmetry is vertical then $x_1=x_3=x_4=0$. It 
follows from equation (\ref{eq:x1}) that in such case $x_5=0$. This   
implies $\tilde{X}=0$ which finishes the proof.
\bl~\\
Any symmetry $\tilde{X}$ of $\Pen$ generates a 
Killing symmetry $X$ of the metric $g$ on the quotient 4-manifold 
${\cal M}$ of leaves of the foliation $\{{\cal S}_x\}$. 
Moreover, the Lie algebra of symmetries  
$\{\tilde{X}_i\}$ is isomorphic to the algebra $\{X_i\}$ of 
the corresponding Killing symmetries. 
\el
Proof.~\\
We write $\tilde{X}$ in its general form (\ref{eq:ssym}). 
Using the conditions (\ref{eq:x1})-(\ref{eq:x5}) we find $e(x_i)$. The 
differentials (\ref{eq:sys1}) imply the form of commutators $[e,l],~[e,f],$ 
etc. Combining this with $e(x_i)$ we find that 
$[e,\tilde{X}]=Ue+U'\bar{e}$, where $U, U'$ are certain complex functions 
on $\Pen$. This implies that vectors of $\tilde{X}$ calculated at points of 
the same leaf of the foliation $\cal V$ differ by a vertical 
part $V=U''e+\bar{U}''\bar{e}$ only. Thus $\tilde{X}$ uniquely projects to 
$X$ on $\cal M$.  $X$ is not zero since any nonzero symmetry $\tilde{X}$ 
has always a nonzero $(f,\bar{f},t,l)$ part. Consider now two nonzero 
symmetries $\tilde{X}=X+x_5e+\bar{x}_5\bar{e}$ 
and $\tilde{Y}=Y+y_5e+\bar{y}_5\bar{e}$. Here $X,~Y$ denote  
$(f,\bar{f},t,l)$ parts of $\tilde{X}$ and $\tilde{Y}$, respectively. 
The commutator of $\tilde{X}$ and $\tilde{Y}$ has the form\\

\noindent
$[\tilde{X},\tilde{Y}]=[X,Y]+x_5[e,Y]+\bar{x}_5[\bar{e},Y]-
y_5[e,X]+\bar{y}_5[\bar{e},X]+(x_5\bar{y}_5-\bar{x}_5y_5)[e,\bar{e}]$ 
modulo $e$ and $\bar{e}$.\\ 

\noindent
Since for any symmetry $\tilde{Z}$ we have $[e,Z]=0$ 
modulo $e$ and $\bar{e}$, then  
$[\tilde{X},\tilde{Y}]=[X,Y]+(x_5\bar{y}_5-\bar{x}_5y_5)[e,\bar{e}]$ 
modulo $e$ and $\bar{e}$. It follows from equations 
(\ref{eq:sys1}), (\ref{eq:domy1}) that $[e,\bar{e}]=0$. Thus  
$[\tilde{X},\tilde{Y}]=[X,Y]$ modulo terms which vanish under 
the projection $\pi:\Pen\to \cal M$. Thus the Lie algebra of 
$\{\tilde{X}_i\}$ is the same as that of $\{X_i\}$.\\
Finally we note that if $\tilde{X}=X+x_5e+\bar{x}_5\bar{e}$ then   
the symmetry equations (\ref{eq:sym2}) and the differentials 
(\ref{eq:sys1}), (\ref{eq:domy2}) imply that  
$$
{\cal L}_X\La=-x_5\bar{F}-\bar{x}_5F,
$$
$$
{\cal L}_XF=-x_5T+\bar{x}_5(3\bar{\al}+\la)\La,
$$
$$  
{\cal L}_XT=\bar{x}_5(3\bar{\al}+\la)\bar{F}+x_5(3\al+\la)F.
$$
These equations imply that on $\Pen$ we have ${\cal L}_XG=0$. This equation 
projects to the equation ${\cal L}_X g=0$ on $\cal M$, since both $X$ and
$G$ have unique projections to $X$ and $g$ on $\cal M$, respectively. 
This finishes the proof of the Lemma.\\

\noindent
\bd~\\
A generalized Einstein bundle of null direction is called (locally) 
homogeneous 
if it posesses six symmetries, which generate a (local) transitive group of 
transformations of $\Pen$.
\ed
It is clear from Lemma 3
that on homogeneous generalized Einstein bundles of null directions all the 
Cartan invariant functions are constant. In such a case we may 
interpret equations (\ref{eq:sys1}) as the Cartan structure equations 
for the left invariant forms on a certain Lie group. This shows that 
(locally) homogeneous generalized 
Einstein bundles of null directions are (local) Lie 
groups whose structure constants may be read from the equations 
(\ref{eq:sys1}), (\ref{eq:domy2}). To determine all the possible groups  
that are homogeneous generalized 
Einstein bundles of null directions we need to check  
which constants $\al,\g_1,\g_4,\om_1,\om_2,\om_3,\om_4$ are compatible with 
the equations (\ref{eq:sys1}), (\ref{eq:domy1})-(\ref{eq:domy2}). 
We find that if  
$\al, \g_1, \g_3,\om_1,\om_2,\om_3,\om_4$ are constants then     
the equations (\ref{eq:domy1}) imply that 
$a=\al_1=\al_4=a_1=a_4=h_1=h_4=\g_1=\g_4=0$ and $h=-(3\al+\la)^2$. Moreover, 
the combination of equations (\ref{eq:domy1}) and (\ref{eq:sys1}) leads to 
two possibilities only.
\begin{itemize}
\item[($i$)] $\al=\la=h=a=0$, $\Om=\om_3T$, $\G=0$. 
\be
\label{eq:64}
\ee
\item[($ii$)] $\al=-\frac{1}{2}\la$, $h=-\frac{1}{4}\la^2$, $a=0$, $\Om=0$, 
$\G=\frac{1}{2}\la E$. 
\end{itemize}
Thus we have two families of homogeneous Einstein bundles. To characterize 
the corresponding groups we note that the family ($i$) leads to the 
vacuum ($\la=0$) Einstein 4-manifold $\cal M$ which has a metric of 
Petrov type $N$ (compare with (\ref{eq:N})). The family ($ii$) leads to  
a 4-metric of Petrov type D (compare (\ref{eq:D})) 
satisfying the Einstein equations with cosmological constant $\la$. The following two examples deal 
explicitly with cases ($i$) and ($ii$). They correspond to 
all possible conformally non-flat solutions to the 
vacuum Einstein equations (with or without cosmological constant) that have 
a 6-dimensional group of symmetries.\\

\noindent
{\bf Example 6} (Example 2 continued)\\
Since Example 2 corresponds to all possible vacuum Einstein spaces with 6 
symmetries, it therefore also exhausts all the possibilities for homogeneous 
vacuum generalized Einstein bundles of null directions $\Pen$. The simplest 
among 
them is the bundle $\Pen$ of Example 1. It may be identified 
with a 6-dimensional group, say $G_0$. Interpreting the 
forms $(F,\bar{F}, T,\La,E,\bar{E})$ as the left invariant forms on $G_0$ 
we easily read the structure constants for $G_0$ from the equations
$$
\der F-E\dz T=0
$$
$$
\der T=0
$$
$$
\der \La -\bar{E}\dz F-E\dz\bar{F}=0
$$
$$
\der E-\bar{F}\dz T=0
$$
obtained from (\ref{eq:sys1}) after insertion of 
conditions $(i)$ with $\om_3=0$.\\
Analysis of the structure constants shows that $G_0$, therefore 
$\Pen$, is a 6-dimensional solvable group. This group is isomorphic to the 
group of symmetries of the corresponding plane wave.\\

\noindent
{\bf Example 7}\\
In case $(ii)$ $\Pen$ also may be identified with a 6-dimensional 
group, say $G_\la$, and its forms  
$(F,\bar{F}, T,\La,E,\bar{E})$ can be identified with the left 
invariant forms on $G_\la$. To 
be more explicit we insert conditions $(ii)$ to (\ref{eq:sys1}) obtaining
$$
\der F-E\dz T-\frac{\la}{2}\bar{E}\dz\La=0
$$
$$
\der T-\frac{\la}{2}(E\dz F+\bar{E}\dz \bar{F})=0
$$
\be
\der \La-\bar{E}\dz F-E\dz\bar{F}=0\label{eq:ex7}
\ee
$$
\der E-\bar{F}\dz T+\frac{\la}{2}\La\dz F=0.
$$
Since $\la=0$ corresponds to the equations discussed in the previous example 
we assume that $\la\neq 0$. It is convenient to introduce real 1-forms 
$(A_1, A_2, A_3, A_1', A_2', A_3')$ defined by 
$$
F=-\frac{1}{\sqrt{2|\la|}}(A_1+i A_1')
$$
$$
T=\frac{1}{2}(A_2-A_2')
$$
\be
\La=\frac{1}{\la}(A_2+A_2')\label{eq:ex71}
\ee
$$
E=\frac{1}{\sqrt{2|\la|}}(A_3+iA_3').
$$
Equations (\ref{eq:ex7}) written in terms of these forms then become 
$$\der A_1=A_2\dz A_3\quad\quad\quad\quad\quad\der A_1'=-A_2'\dz A_3'$$
\be
\der A_3=-A_1\dz A_2\quad\quad\quad\quad\quad\der A_3'=-A_1'\dz A_2'
\label{eq:ex72}
\ee
$$
\der A_2=-\eps A_3\dz A_1\quad\quad\quad\quad\quad\der A_2'=-\eps A_3'\dz A_1',
$$
where $\eps=\pm 1=$(sign of $\la$). These equations show that the 
group $G_\la$ is a direct product of two groups $H$ and $H'$. The Lie algebra 
of $H$ is isomorphic to ${\bf sl}(2,{\bf R})$ and the Lie algebra of $H'$  
depends on the sign of $\la$ and is isomorphic to ${\bf su}(2)$ if $\la>0$ 
and to ${\bf sl}(2,{\bf R})$ if $\la <0$. Thus in this case 
$\Pen\equiv G_\la\equiv H\times H'$. In terms of the variables $A_i$, $A_i'$ the 
degenerate metric $G$ of Theorem 1 has the form 
$G= \frac{1}{\la}[A_1^2-\eps A_2^2+A_1'^2+\eps A_2'^2]$. According to this 
theorem the space of 
leaves of the fibration is equipped with the Einstein metric $g$ with 
nonvanishing cosmological constant $\la$. \\
$(ii~a)$ Assume that $\la>0$.\\
In this case the Lie algebra of $H$ is isomorphic to ${\bf sl}(2,{\bf R})$ 
and the Lie algebra of $H'$ is isomorphic to ${\bf su}(2)$. In the following 
we concentrate on the case when $H={\bf SO}(1,2)$ and $H'={\bf SO}(3)$, but 
one can also consider cases in which $H$ and/or $H'$ are double covers of 
these groups.\\
If $H={\bf SO}(1,2)$ and $H'={\bf SO}(3)$ then a 
coordinate system $(x_1,x_2,x_3)$ on $H$ and 
a coordinate system $(x_1',x_2',x_3')$ on $H'$ may be chosen such that
$$
A_1=\cosh x_2 \cosh x_3 \der x_1-\sinh x_3\der x_2
$$
$$
A_2=-\cosh x_2 \sinh x_3 \der x_1+\cosh x_3\der x_2
$$
\be
A_3=\sinh x_2 \der x_1+\der x_3
\label{eq:ex74}
\ee
$$
A_1'=\cos x_2' \cos x_3' \der x_1'+\sin x_3'\der x_2'
$$
$$
A_2'=-\cos x_2' \sin x_3' \der x_1'+\cos x_3'\der x_2'
$$
$$
A_3'=\sin x_2' \der x_1'+\der x_3'.
$$
Since the above forms satisfy equations (\ref{eq:ex72}) for $\eps=1$, 
then for each value of $\la>0$, via (\ref{eq:ex71}), they define a solution 
to the system (\ref{eq:sys1}) with $\al,\Om,\G$ given by (\ref{eq:64}) 
$(ii)$.\\
To obtain a better insight into this solution and its corresponding space-time 
consider a generic point $P\in\Pen={\bf SO}(1,2)\times{\bf SO}(3)$ which in 
coordinates $(x_1,x_2,x_3,x_1',x_2',x_3')$ can be represented by 
a $6\times 6$ matrix of the form 
$$
P=\left(\begin{array}{cc}
p&0\\
0&p'
\end{array}\right),\label{eq:zona}
$$
where $p=p_1p_2p_3$, $p'=p_1'p_2'p_3'$ and the 1-parameter groups $p_i$, $p_i'$
are given by 
$$
\begin{array}{cc}
p_1 =\left(\begin{array}{ccc}
1&0&0\\
0&\cos x_1&-\sin x_1\\
0&\sin x_1&\cos x_1
\end{array}\right),
&
p_2 =\left(\begin{array}{ccc}
\cosh x_2&0&-\sinh x_2\\
0&1&0\\
-\sinh x_2&0&\cosh x_2
\end{array}\right),
\end{array}
$$
$$
\begin{array}{cc}
p_3 =\left(\begin{array}{ccc}
\cosh x_3&\sinh x_3&0\\
\sinh x_3&\cosh x_3&0\\
0&0&1
\end{array}\right),
&
p_1' =\left(\begin{array}{ccc}
1&0&0\\
0&\cos x_1'&-\sin x_1'\\
0&\sin x_1'&\cos x_1'
\end{array}\right),
\end{array}
$$
$$
\begin{array}{cc}
p_2' =\left(\begin{array}{ccc}
\cos x_2'&0&\sin x_2'\\
0&1&0\\
-\sin x_2'&0&\cos x_2'
\end{array}\right),
&
p_3' =\left(\begin{array}{ccc}
\cos x_3'&-\sin x_3'&0\\
\sin x_3'&\cos x_3'&0\\
0&0&1
\end{array}\right).
\end{array}
$$
Then the forms $A_i$, $A_i'$ can be identified with the components of the 
Maurer-Cartan form $A=P^{-1}\der P$ on $\Pen$ by
$$
A=P^{-1}\der P=\left(\begin{array}{cccccc}
0&A_3&-A_2&&&\\
A_3&0&-A_1&&&\\
-A_2&A_1&0&&&\\
&&&0&-A_3'&A_2'\\
&&&A_3'&0&-A_1'\\
&&&-A_2'&A_1'&0
\end{array}\right).
$$
Consider now the degenerate metric 
$G=\frac{1}{\la}[A_1^2-A_2^2+A_1'^2+A_2'^2]$ on $\Pen$ and 
a subgroup ${\bf SO}(1,1)\times{\bf SO}(2)$ of 
${\bf SO}(1,2)\times{\bf SO}(3)$ given by those elements which have the form  
$$
g_*=\left(\begin{array}{cccccc}
\cosh t_3&\sinh t_3&0&&&\\
\sinh t_3&\cosh t_3&0&&&\\
0&0&1&&&\\
&&&\cos t_3'&-\sin t_3'&0\\
&&&\sin t_3'&\cos t_3'&0\\
&&&0&0&1
\end{array}\right),
$$
where $t_3\in{\bf R}$, $t_3'\in [0,2\pi]$. It follows that the left 
action ($P\to g_*P$) of this group on 
$\Pen={\bf SO}(1,2)\times{\bf SO}(3)$ leaves the form 
$A$ invariant. The right action ($P\to Pg_*$) transforms $A$ according to 
$A\to g_*^{-1}Ag_*$. These relations imply that the metric $G$ is invariant under 
the right action so that it projects to the homogeneous space 
${\cal M}={\bf SO}(1,2)\times{\bf SO}(3)/{\bf SO}(1,1)\times{\bf SO}(2)$ 
equipped with the Einstein ($\la>0)$ metric 
$$g=\frac{1}{\la}[\cosh^2x_2\der x_1^2-\der x_2^2+\cos^2x_2'\der x_1'^2
+\der x_2'^2].
$$ 
This shows that $\Pen={\bf SO}(1,2)\times{\bf SO}(3)$ is fibred over the 
Einstein space-time ${\cal M}={\bf H}_{+-}\times{\bf S}^2$, which is a 
Cartesian product of a neutral-signature hyperbolic space
and a 2-sphere, both with their natural metrics
\footnote{${\bf H}_{+-}=
\{(z_1,z_2,z_3)\in {\bf R}^{1,2}~|~-z_1^2+z_2^2+z_3^2=1\}$, a quadric in 
${\bf R}^3$ equipped with the flat Lorentzian metric of signature 
(-,+,+). A parametrization of ${\bf H}_{+-}$ used above is: 
$z_1=\sinh x_2$, $z_2=\cosh x_2\sin x_1$, $z_3=\cosh x_2\cos x_1$. 
The sphere is parametrized by $z_1=\cos x_1'\cos x_2'$, 
$z_2=\sin x_1'\cos x_2'$ and $z_3'=\sin x_2'$.}. It is clear that the fibres 
of $\Pen$, being homeomorphic to ${\bf SO}(1,1)\times{\bf SO}(2)$ have the  
topology of a cylinder. \\\\

\noindent
$(ii~b)$ Assume that $\la<0$.\\
Now the Lie algebras of both $H$ and $H'$ are isomorphic to 
${\bf sl}(2,{\bf R})$. We again 
concentrate on the case when $H={\bf SO}(1,2)$ and $H'={\bf SO}(1,2)$. 
Introducing $(x_1,x_2,x_3,x_1',x_2',x_3')$ as local coordinates on 
$\Pen=H\times H'$ we can represent any point $P$ of $\Pen$ as
$$
P=\left(\begin{array}{cc}
p&0\\
0&p'
\end{array}\right),
$$
where $p=p_1p_2p_3$, $p'=p_1'p_2'p_3'$ and the 1-parameter groups $p_i$, $p_i'$
are given by 
$$
\begin{array}{cc}
p_1 =\left(\begin{array}{ccc}
1&0&0\\
0&\cosh x_1&\sinh x_1\\
0&\sinh x_1&\cosh x_1
\end{array}\right),
&
p_2 =\left(\begin{array}{ccc}
\cosh x_2&0&-\sinh x_2\\
0&1&0\\
-\sinh x_2&0&\cosh x_2
\end{array}\right),
\end{array}
$$
$$
\begin{array}{cc}
p_3 =\left(\begin{array}{ccc}
\cos x_3&\sin x_3&0\\
-\sin x_3&\cos x_3&0\\
0&0&1
\end{array}\right),
&
p_1' =\left(\begin{array}{ccc}
1&0&0\\
0&\cosh x_1'&\sinh x_1'\\
0&\sinh x_1'&\cosh x_1'
\end{array}\right),
\end{array}
$$
$$
\begin{array}{cc}
p_2' =\left(\begin{array}{ccc}
\cos x_2'&0&\sin x_2'\\
0&1&0\\
-\sin x_2'&0&\cos x_2'
\end{array}\right),
&
p_3' =\left(\begin{array}{ccc}
\cosh x_3'&-\sinh x_3'&0\\
-\sinh x_3'&\cosh x_3'&0\\
0&0&1
\end{array}\right).
\end{array}
$$
Then the Maurer-Cartan form $A=P^{-1}\der P$ defines forms $A_i$ and $A_i'$ 
by 
$$
A=P^{-1}\der P=\left(\begin{array}{cccccc}
0&A_3&-A_2&&&\\
-A_3&0&A_1&&&\\
-A_2&A_1&0&&&\\
&&&0&-A_3'&A_2'\\
&&&-A_3'&0&A_1'\\
&&&-A_2'&A_1'&0
\end{array}\right).
$$ 
The subgroup ${\bf SO}(2)\times{\bf SO}(1,1)$ of 
${\bf SO}(1,2)\times{\bf SO}(1,2)$ consisting of elements $g_*$ of the form 
$$
g_*=\left(\begin{array}{cccccc}
\cos t_3&\sin t_3&0&&&\\
-\sin t_3&\cos t_3&0&&&\\
0&0&1&&&\\
&&&\cosh t_3'&-\sinh t_3'&0\\
&&&-\sinh t_3'&\cos t_3'&0\\
&&&0&0&1
\end{array}\right),
$$
where $t_3\in [0,2\pi]$, $t_3'\in{\bf R}$, 
acts on $\Pen$ from the right. The form $A$ transforms by $A\to g_*^{-1}Ag_*$ 
under this action which implies that the metric 
$G=\frac{1}{|\la|}[A_1^2+A_2^2+A_1'^2-A_2'^2]$ is invariant. The quotient 
space 
${\cal M}={\bf SO}(1,2)\times {\bf SO}(1,2)/{\bf SO}(2)\times{\bf SO}(1,1)$
is naturally equipped with the projected Einstein ($\la<0$) metric 
$$g=\frac{1}{|\la|}[\cosh^2x_2\der x_1^2+\der x_2^2+\cos^2x_2'\der x_1'^2
-\der x_2'^2].
$$ 
It follows that now ${\cal M}={\bf H}_{++}\times{\bf H}_{+-}$, that is it 
is a Cartesian product of Euclidean- and neutral-signature hyperbolic 
2-spaces
\footnote{${\bf H}_{++}=
\{(z_1,z_2,z_3)\in {\bf R}^{1,2}~|~-z_1^2+z_2^2+z_3^2=-1\}$, a quadric in 
${\bf R}^3$ equipped with flat Lorentzian metric of signature 
(-,+,+). The parametrization of ${\bf H}_{++}$ used here is: 
$z_1=\cosh x_2$, $z_2=\sinh x_2\sin x_1$, $z_3=\sinh x_2\cos x_1$. 
${\bf H}_{+-}$ is parametrized by $z_1=\cos x_2'\sinh x_1'$, 
$z_2=\cos x_2'\cosh x_1'$, $z_3=\sin x_2'$.}. The metric $g$ is a 
sum of the natural metric on ${\bf H}_{++}$ minus the natural metric on 
${\bf H}_{+-}$. The fibres of $\Pen$ 
have again the topology of a cylinder\footnote{Note that the metrics 
$\la^{-1}(-g_{H+-}+g_{S})$, and $|\la|^{-1}(g_{H++}+g_{H+-})$, where 
$g_{H++}$, $g_{H+-}$ and $g_{S}$ denote the natural metrics on 
${\bf H}_{++}$, ${\bf H}_{+-}$ and ${\bf S}^2$ respectively, do not 
satisfy the vacuum Einstein equations with cosmological constant. 
It is interesting to note that the first of these is a solution to the 
Einstein-Maxwell equations with vanishing cosmological constant 
known as the Bertotti-Robinson solutions. The second metric has an  
energy momentum with negative energy and thus can not be 
interpreted as the Einstein-Maxwell solution.}.\\

\noindent
Example 7 shows that if $\la\neq 0$ then the {\it homogeneous} generalized bundles of null 
directions $\Pen$ are principal fibre bundles over the space-time 
with the structure group $G_*={\bf SO}(2)\times{\bf SO}(1,1)$. 
The system of 1-forms of Theorem 1 on these bundles equips them with a 
1-form $A$ which is valued in the Lie algebra of group $G_\la$ such that 
dim$G_\la=$dim$\Pen$. Moreover, $A$ has the following properties.
\begin{itemize}
\item[-] if $X$ is a vector field tangent to the flow of the one parameter 
subgroup of $G_*$ generated by $\xi$ then 
$A(X)=\xi$,
\item[-] $A(X)\neq 0$ on each vector tangent to $\Pen$,
\item[-] under the right action of $G_*$ the form $A$ 
transforms as $g_*^{-1}Ag_*$.
\end{itemize}

\noindent
Thus $A$ can be understood as a Cartan connection on $\Pen$ 
(compare with \cite{bi:Kobayashi} pp. 127-130). Note that in Example 7 
the form $A$ is always a Maurer-Cartan form on $G_\la$ which implies that 
its curvature is  zero. This suggests that a generic (nonhomogeneous) 
generalized Einstein bundle of null directions with nonzero cosmological 
constant can find a useful formulation in terms of curvature 
conditions on Cartan connections on principal fibre bundles with group $G_*$ over the 
space-time. Such Cartan connections on $\Pen$ can be further understood as 
the usual connections on fibre bundles $\Pen\times_{G_*}G$ 
(compare with \cite{bi:Kobayashi}, pp. 127-128). This possibility will be 
studied 
elsewhere.\\

\section{Concluding remarks}

The study of null objects in general relativity has led to many important
advances in the understanding of Einstein's equations.   In this paper
this general line of enquiry has been developed by employing the bundle of
null directions, over a four dimensional Lorentzian space-time, as a tool
in the investigation of Einstein spaces.  It has been shown that a
Lorentzian 4-metric can be defined by a differential system on a six
manifold over a four manifold.  In fact a G-structure on the six manifold
(the total space of the bundle of null directions) encodes the requirement
that the four manifold be Einstein.  This structure, as has been
demonstrated, can be used to study space-times.  An extension of this
structure leads to the construction of a generalised bundle of null
directions over a conformally non-flat Einstein space-time.  An effective
algorithm for the equivalence problem for Lorentizian 4-metrics has been
constructed by making use of this generalised bundle.  Finally it has been
observed that the Petrov type of the Weyl tensor of a conformally non-flat
Einstein metric can be encoded in the fibration of a six manifold over a
four manifold. Different fibrations provide interesting insights into
Einstein space-times.

\end{document}